\documentclass[letterpaper]{article} 

\ifdefined\aaaianonymous
    \usepackage[submission]{aaai2027} 
\else
    \usepackage[preprint]{aaai2027}   
\fi

\usepackage[hyphens]{url} 
\usepackage{graphicx}   
\usepackage{natbib}     
\usepackage{caption}    
\usepackage{amsmath}
\usepackage{amssymb}
\usepackage{booktabs}
\usepackage{multirow}
\usepackage{algorithm}
\usepackage{algorithmic}
\usepackage{xspace}
\usepackage[capitalize]{cleveref}

\newcommand{\forge}{\textsc{Forge}\xspace}
\newcommand{\gepa}{\textsc{Gepa}\xspace}
\newcommand{\miprov}{\textsc{MiproV2}\xspace}
\newcommand{\ace}{\textsc{Ace}\xspace}

\newcommand{\Call}[2]{\textsc{#1}(#2)}

\crefname{algorithm}{Algorithm}{Algorithms}
\Crefname{algorithm}{Algorithm}{Algorithms}

\title{Failure-Guided Co-Evolution of Prompts and Training Data}

\author{Tianyu Yuan, Zhuzhong Qian\textsuperscript{*}}
\affiliations{
    Nanjing University\\
    yty@smail.nju.edu.cn, qzz@nju.edu.cn\\
    \textsuperscript{*}Corresponding author.
}

\begin{document}

\maketitle

\begin{abstract}
Automatic prompt optimization (APO) improves language-model programs by revising prompts from task feedback, yet it typically holds its training data fixed.
Repeatedly optimizing against the same instances confines feedback to weaknesses already represented in those data, leaving related failure conditions unexplored.
We therefore view each failure as a dual signal: it indicates both how the prompt should be revised and what new training evidence should be synthesized.
We introduce \forge, a failure-guided framework that co-evolves prompts and training data.
\forge abstracts imperfect executions into reusable failure modes and synthesizes new training data through four complementary mutation strategies.
Verified instances are fed back into prompt search, allowing updated prompts to expose the next data needs.
Across eight heterogeneous benchmarks, \forge improves the aggregate score over the unoptimized baseline by 16.52 percentage points and outperforms all evaluated APO baselines.
The synthesized data also transfer beyond \forge: in a transfer study, they improve all nine APO comparisons by 2--9 points and all three GRPO comparisons by 4--8 points under matched optimization budgets.
These results establish failures as a shared interface between prompt optimization and data synthesis, and show the benefit of jointly adapting what a model is instructed to do and what it learns from.
\end{abstract}

\section{Introduction}
\label{sec:introduction}

\begin{figure}[t]
    \centering
    \includegraphics[width=\columnwidth]{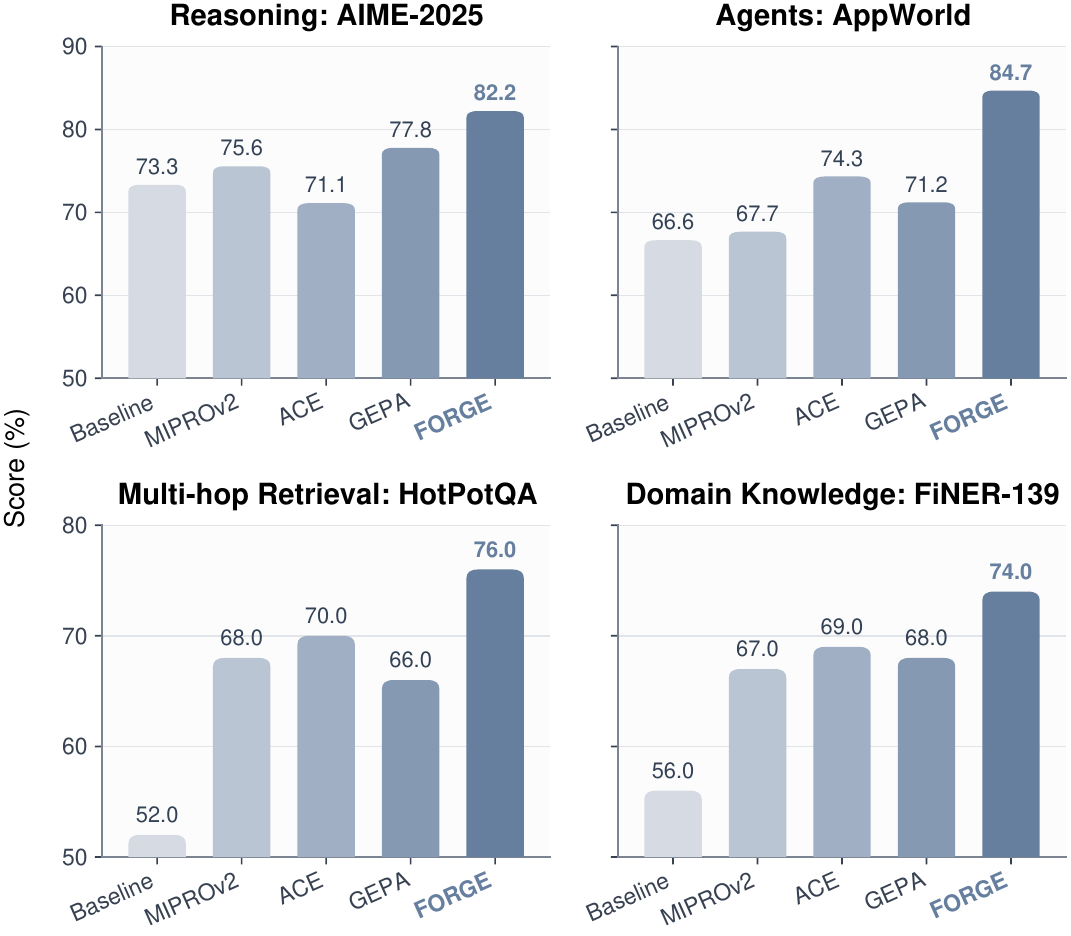}
    \caption{Results on representative benchmarks from four task categories.
    \forge consistently outperforms all compared methods across multi-hop
    retrieval, domain knowledge, reasoning, and agent tasks, demonstrating its
    effectiveness across diverse benchmarks.}
    \label{fig:category-benchmark-results-2x2}
\end{figure}

Language-model (LM) programs combine prompted calls with retrieval, tools, and intermediate computation for applications ranging from structured prediction and knowledge-intensive reasoning to instruction following and interactive agents \cite{opsahlong2024mipro,trivedi2024appworld}.
Because their behavior is largely specified by natural-language prompts, automatic prompt optimization (APO) replaces manual design with feedback-driven search.
Existing methods use textual gradients, optimizer LMs, joint instruction--demonstration search, or evolutionary reflection \cite{pryzant2023protegi,yuksekgonul2025textgrad,yang2024opro,opsahlong2024mipro,fernando2024promptbreeder,agrawal2026gepa}.

Despite these advances, most general-purpose APO systems still derive feedback from fixed training data \cite{opsahlong2024mipro,agrawal2026gepa,zhang2026ace}.
As prompts evolve, repeatedly optimizing on the same instances can overemphasize observed failures while leaving nearby conditions unexplored.
When performance stops improving, it is therefore unclear whether prompt search has exhausted useful revisions or the training data have stopped exposing informative failures.
This motivates treating training data as an adaptive optimization state.

Feedback-directed synthesis has been studied mainly for model training: early methods bootstrap from seed tasks or model priors \cite{wang2023selfinstruct,xu2025magpie}, whereas learner-aware methods generate instances from observed errors or task-model feedback \cite{lee2024llm2llm,menon2024discern,li2025reversegen,khan2025dataenvgym}.

Recent work couples data with prompt or model updates: a human-in-the-loop workflow jointly revises prompts and a living test set \cite{lee2026dataprompt}, while \textsc{CoEvolve} synthesizes interaction tasks during weight optimization \cite{yang2026coevolve}.
Together, they motivate a complementary question: can automated reflective search over discrete prompts for heterogeneous LM programs treat training data as a mutable optimization state?
We also ask whether data produced inside this loop remain useful for subsequent LLM weight optimization.

\begin{figure*}[t]
    \centering
    \includegraphics[width=\textwidth]{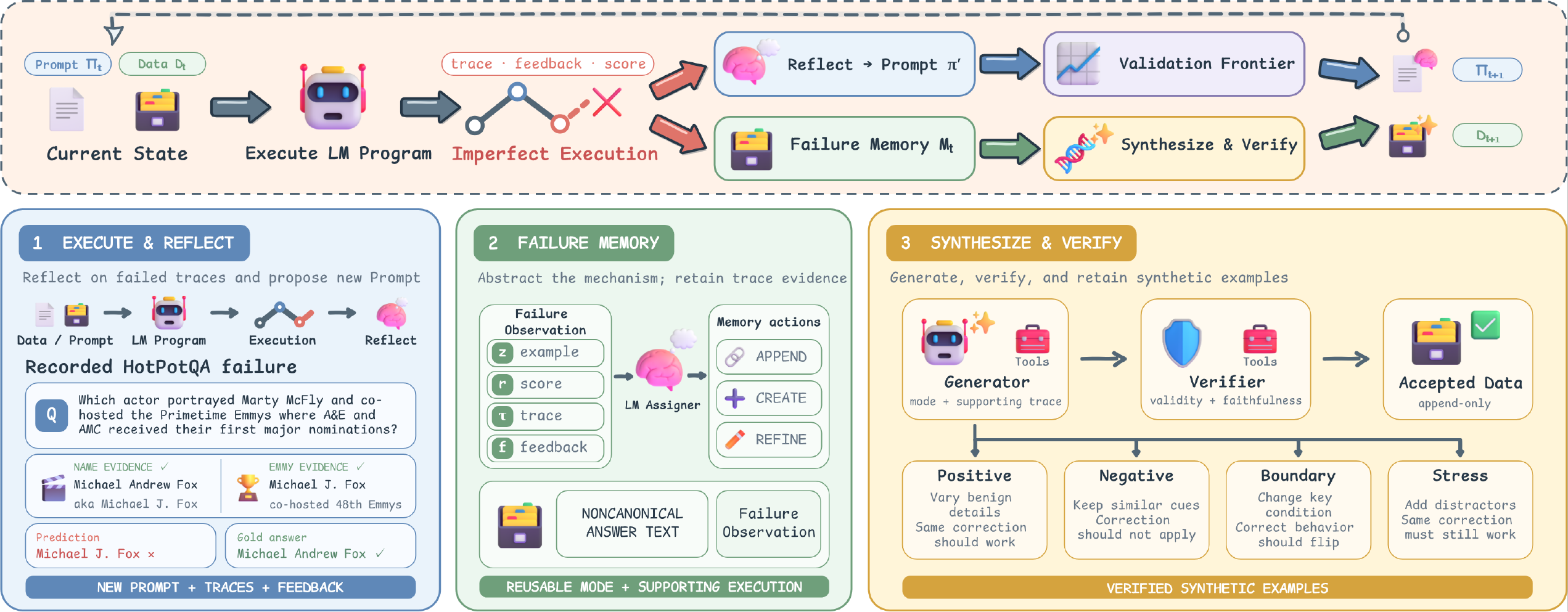}
    \caption{Overview of \forge.
    Execution failures drive prompt reflection and update failure memory; when aggregate validation performance stops improving, active modes guide four verified mutations whose accepted instances re-enter training.}
    \label{fig:overview}
\end{figure*}

We address this problem with \forge, a failure-guided framework that co-evolves prompts and training data.
Its key insight is that each failure reveals both how to revise the prompt and what training evidence is missing.
\forge uses failed executions for prompt reflection and abstracts their recurring weaknesses into a failure memory.
When the best aggregate validation score stops improving, active failure modes guide data synthesis, completing the prompt--data loop in \Cref{fig:overview}.

\forge expands active failure modes through four mutation intents: \emph{positive} preserves the target skill under benign variation, \emph{negative} creates contrasts where the failed heuristic should not apply, \emph{boundary} probes nearby decision conditions, and \emph{stress} adds distractors or interacting constraints.
A separate tool-capable verifier agent admits only valid instances faithful to both the targeted failure mode and mutation intent.
Accepted instances enter the training data and shape subsequent prompt search, whose failures refresh the memory and guide the next synthesis event.

We evaluate \forge on prompt optimization and data transfer; \Cref{fig:category-benchmark-results-2x2} shows representative results across four task categories.
Across eight benchmarks, \forge raises average performance by \textbf{16.52} points over the unoptimized baseline and by \textbf{4.92--8.82} points over competing APO methods.
In a separate transfer study, its synthesized instances improve all nine prompt-optimizer and all three GRPO comparisons, including \textbf{4--8}-point GRPO gains, demonstrating utility beyond \forge.

Our contributions are fourfold:

\begin{itemize}
    \item We treat training data as an adaptive state of APO and formulate a failure-mediated loop that co-evolves prompt candidates and training data.
    \item We introduce failure-mode-guided synthesis with positive, negative, boundary, and stress mutations.
    \item We evaluate \forge against strong APO baselines across eight heterogeneous LM-program benchmarks.
    \item We use a paired GRPO study to show that \forge-generated data transfer to weight optimization and improve final performance under matched optimization budgets.
\end{itemize}

\section{Related Work}
\label{sec:related-work}

\paragraph{Automatic prompt and context optimization.}

Automatic prompt optimization (APO) uses task feedback to search natural-language instructions and contextual controls.
Representative methods translate errors into language feedback (\textsc{ProTeGi}, \textsc{TextGrad}), condition optimizer LMs on scored histories (\textsc{OPRO}), or evolve and revise prompts through mutations and failure patterns (\textsc{PromptBreeder}, \textsc{AMPO}) \cite{pryzant2023protegi,yuksekgonul2025textgrad,yang2024opro,fernando2024promptbreeder,yang2024ampo}.
For multistage LM programs, \miprov searches instructions and demonstrations, \gepa combines trace reflection with Pareto selection, \textsc{AutoPDL} uses successive halving, and \ace curates a structured playbook \cite{opsahlong2024mipro,agrawal2026gepa,spiess2025autopdl,zhang2026ace}.
Despite their different search strategies, these methods keep training data fixed; \forge instead allows failures to expand the data that drive subsequent search.

\paragraph{Data-augmented prompt optimization.}
Recent systems adapt examples or evaluation artifacts alongside prompts: PromptWizard refines instructions and synthetic demonstrations, Promptomatix synthesizes task-specific data from descriptions, and Data-Prompt Co-Evolution jointly revises prompts and a living test set with developers \cite{agarwal2025promptwizard,murthy2025promptomatix,lee2026dataprompt}.
SIPDO organizes synthesis with a controlled difficulty tier and progressively challenges the prompt using generated examples \cite{yu2026sipdo}.
CASPER optimizes prompt representations in a continuous embedding space using textual feedback and augments optimization with failure cases \cite{jain2026casper}.
\forge instead turns recurring failures into tool-grounded synthesis and verification targets for heterogeneous LM programs, then evaluates the resulting instances under other prompt optimizers and GRPO.

\paragraph{Learner-aware training-data synthesis.}
Self-Instruct, Magpie, CodecLM, Evol-Instruct, and RandomWorld bootstrap, mutate, or procedurally generate instruction and tool-use data without adapting to learner failures \cite{wang2023selfinstruct,xu2025magpie,wang2024codeclm,xu2024wizardlm,sullivan2025randomworld}.
Learner-aware methods instead expand mispredictions (\textsc{LLM2LLM}), summarize error clusters (\textsc{DISCERN}), select or synthesize from weakness profiles (\textsc{DataEnvGym}, \textsc{STAT}), or search for failure-inducing queries (\textsc{ReverseGen}) \cite{lee2024llm2llm,menon2024discern,khan2025dataenvgym,he2026stat,li2025reversegen}.
Executable self-play further couples synthesis with weight updates by validating generated tasks through code execution (\textsc{Absolute Zero}) or environment interaction (\textsc{CoEvolve}) \cite{zhao2025absolutezero,yang2026coevolve}.
\forge instead uses execution failures to couple discrete prompt reflection with data synthesis across heterogeneous LM programs, then evaluates whether its instances transfer to other prompt optimizers and GRPO-based weight optimization.

\section{Problem Formulation}
\label{sec:formulation}

\paragraph{LM-program prompt optimization.}
Let $F_{\boldsymbol{\pi}}$ denote an LM program whose underlying model weights are frozen and whose $M$ optimizable module prompts are collected in $\boldsymbol{\pi}=(\pi^{(1)},\ldots,\pi^{(M)})$.
A task instance $z\in\mathcal Z$ may represent either a conventional input or an interaction with a task environment.
Executing $F_{\boldsymbol{\pi}}$ on $z$ produces an ordered message trace $\tau$, task feedback $f$, and a normalized score $r(F_{\boldsymbol{\pi}},z)\in[0,1]$; we call the execution imperfect when $r(F_{\boldsymbol{\pi}},z)<1$.

\paragraph{Fixed-data APO.}
Given initial training data $D_0$, a validation set $V$, and a test set $T$, an APO algorithm draws update minibatches from $D_0$.
Let $\Pi_K$ contain the prompt candidates it has evaluated by stopping time $K$.
The final prompt is selected by the fixed validation objective
\begin{equation}
    J_V(\boldsymbol{\pi})=\frac{1}{|V|}\sum_{z\in V}r(F_{\boldsymbol{\pi}},z),
    \qquad
    \widehat{\boldsymbol{\pi}}=\arg\max_{\boldsymbol{\pi}\in\Pi_K}J_V(\boldsymbol{\pi}).
    \label{eq:apo-objective}
\end{equation}
The validation set may control search and candidate selection but never provides training or synthesis instances; test scores never affect either process and are used only for reporting.
Conventional fixed-data APO additionally imposes $D_t=D_0$ throughout search.

\paragraph{Adaptive-data APO.}
We retain the same validation objective while allowing training data to evolve.
At iteration $t$, let $\Pi_t$ be the candidate pool, $D_t$ the training multiset, and $\mathcal H_t$ the history of imperfect-execution observations $o=(z,\boldsymbol{\pi},\tau,f,r)$.
Abstracting away the particular search and synthesis mechanisms, the coupled updates are
\begin{equation}
    \begin{aligned}
        (\Pi_{t+1},\mathcal H_{t+1})
        &=\mathcal U(\Pi_t,\mathcal H_t;D_t),\\
        D_{t+1}&=D_t\uplus\mathcal A(\mathcal H_{t+1}).
    \end{aligned}
    \label{eq:adaptive-apo}
\end{equation}
where $\mathcal U$ denotes a prompt-search update over the current training data, $\mathcal A$ returns accepted synthetic instances, and $\uplus$ denotes multiset addition.
This update is append-only: accepted synthetic instances may be added, while existing training instances are neither removed nor modified.
Fixed-data APO is the special case $\mathcal A\equiv\varnothing$.
Thus data synthesis changes the evidence available to subsequent prompt updates without changing the validation objective.

\section{Method}
\label{sec:method}

\forge couples prompt optimization and training-data synthesis through execution failures while maintaining prompt candidates $\Pi_t$, mutable training data $D_t$, and failure memory $\mathcal M_t$.
Each imperfect execution both guides an immediate prompt revision and becomes a reusable failure mode for later synthesis.
Prompt search thus changes program behavior, while synthesis changes the evidence for subsequent updates (\Cref{fig:overview}).

\subsection{Optimization Loop}
\label{sec:optimization-loop}

At iteration $t$, \forge selects a parent prompt $\boldsymbol{\pi}\in\Pi_t$, samples a minibatch $B\subset D_t$, and executes the LM program to collect scores, ordered traces, and task feedback.
Each imperfect execution then branches in two directions.
The prompt branch reflects on the observed trace and feedback to propose a revised prompt $\boldsymbol{\pi}'$; the memory branch uses the same evidence to update failure memory, as described in the next subsection.
Using the same failures in both branches keeps prompt revision and data synthesis focused on weaknesses actually exposed by the current program.

The prompt branch uses an independent reflective-search backbone inspired by \gepa \cite{agrawal2026gepa}.
The parent and proposal are compared on the same minibatch, and the proposal enters the candidate pool only when it improves the aggregate minibatch score.
Each admitted candidate is then evaluated on the fixed validation set: per-instance scores support frontier-based parent selection, while aggregate scores govern synthesis scheduling and final prompt selection.

The data branch becomes active after $p$ completed prompt-search steps without a strict improvement in the best aggregate validation score observed so far.
\forge uses the active failure memory to synthesize a budget of new instances, admits only verified instances, and appends them to the training data to obtain $D_{t+1}$.
Prompt search then continues on the expanded data.
The new instances can expose failures absent from $D_t$; those failures in turn revise later prompts and determine the next synthesis targets.
This closes the optimization loop.
\Cref{alg:forge} summarizes the procedure; further algorithmic and implementation details are provided in Appendix~\ref{app:algorithm}.

\begin{algorithm}[!t]
\caption{\forge: failure-guided prompt and data optimization}
\label{alg:forge}
\begin{algorithmic}[1]
\REQUIRE training data $D$, validation set $V$, seed prompt $\boldsymbol{\pi}_0$, patience $p$, synthesis budget $K$, stopping rule $\mathcal S$
\STATE $\Pi\gets[\boldsymbol{\pi}_0]$; evaluate $\boldsymbol{\pi}_0$ on $V$ and store scores $S_V$; $\mathcal M\gets\varnothing$
\WHILE{$\mathcal S$ is not met}
    \STATE $\boldsymbol{\pi}\gets\Call{SelectCandidate}{\Pi,S_V}$
    \STATE $B\gets\Call{SampleBatch}{D}$
    \STATE $E\gets\Call{Evaluate}{\boldsymbol{\pi},B}$; $E^-\gets\{e_i\in E:r_i<1\}$
    \STATE $\mathcal M\gets\Call{UpdateFailureMemory}{\mathcal M,E^-}$
    \STATE $\boldsymbol{\pi}'\gets\Call{UpdatePrompt}{\boldsymbol{\pi},E^-}$
    \STATE $E'\gets\Call{Evaluate}{\boldsymbol{\pi}',B}$
    \IF{$R_B(\boldsymbol{\pi}')>R_B(\boldsymbol{\pi})$}
        \STATE append $\boldsymbol{\pi}'$ to $\Pi$; evaluate it on $V$ and update $S_V$
    \ENDIF
    \IF{no strict aggregate-validation gain for $p$ steps}
        \STATE $D\gets D\uplus\Call{Synthesize}{\mathcal M,K}$
    \ENDIF
\ENDWHILE
\RETURN best prompt on $V$ and augmented $D$
\end{algorithmic}
\end{algorithm}

\subsection{Failure Memory}
\label{sec:failure-memory}

Directly asking an LM to synthesize from a single failed instance can easily collapse into near-duplicate generation: the model preserves the same entities, input structure, and reasoning pattern while changing only surface wording.
Such variants add little new evidence for prompt optimization because they restate a condition already represented in the training data instead of probing when the failed behavior should or should not occur.
The central challenge is therefore to separate the reusable failure mechanism from the particular instance that exposed it.

\forge addresses this challenge with a failure memory whose entries are \emph{failure modes}: reusable mechanism descriptions paired with the concrete failed executions that support them.
For every imperfect execution, an LM-based assigner examines the task instance, trace, and feedback.
It then assigns the execution to an existing mode, creates a new mode, or refines an existing one.
The mode description abstracts away instance-specific entities, while its supporting executions retain the task context and trace evidence needed to ground later generation.
This separation enables synthesis to vary the conditions around a weakness rather than merely paraphrasing the source failure.

Two failures observed during an AppWorld optimization run \cite{trivedi2024appworld} show why modes are more useful than task-specific records.
Asked to follow every classical Spotify artist with at least 16 followers, the agent followed five artists but missed a sixth because it inspected only the default result page.
The same mode later absorbed a Venmo failure: when asked to accept all pending requests from roommates and coworkers, the agent accepted only two of ten qualifying requests after again processing an incomplete result set.
\forge records both executions as \emph{incomplete paginated search}: the agent mistakes the default result page for the complete candidate set, so qualifying records on later pages are never considered or acted upon.
This abstraction discards app-specific entities and actions while preserving the missing operation, allowing the mode to guide synthesis across unrelated tasks.

The modes accumulated since the preceding synthesis event form the active memory used by the next event.
This keeps synthesis aligned with weaknesses exposed by the current prompt--data state: when later prompts and added data reveal a different mechanism, the new failures induce different modes rather than forcing generation to revisit the same source instances.

\subsection{Failure-Guided Data Synthesis}
\label{sec:synthesis}

When synthesis is triggered, \forge allocates its generation budget across active failure modes, prioritizing modes supported by more observations.
Each generation attempt is conditioned on both a mode description and one of its supporting failed executions, including the trace and feedback.
The mode specifies \emph{what weakness} to probe, while the execution anchors the generated instance in a concrete task setting.
This differs from asking an LM to produce generically difficult data or merely paraphrase a failed input.

A failure mode identifies what behavior is wrong, but not where its correction should and should not generalize.
If synthesis produces only positive variants, the optimizer may overgeneralize the apparent fix or learn a brittle rule tied to the source instance.
\forge therefore constructs a local behavioral neighborhood around each failure mode: positive mutations test whether the correction transfers across new realizations, negative mutations mark where it should not apply, boundary mutations isolate the condition that changes the desired behavior, and stress mutations test whether the correction remains effective under distractors.
Together, these four directions define the scope and robustness of a correction rather than merely diversifying its surface form.

A recorded execution from our HotPotQA optimization run makes these roles concrete.
The program is asked which actor portrayed Marty McFly in the \emph{Back to the Future} trilogy and co-hosted the Primetime Emmy ceremony at which A\&E and AMC received their first major nominations.
The retrieved evidence states that Michael Andrew Fox is known professionally as Michael J. Fox and that Michael J. Fox co-hosted the 48th Primetime Emmy Awards.
The program therefore identifies the correct person but returns \emph{Michael J. Fox}, whereas the benchmark gold answer is \emph{Michael Andrew Fox}.
Failure memory records this execution as \emph{noncanonical answer text}: the reasoning chain reaches the correct entity, but the returned alias does not match the required answer form.
From this recorded failure, the same run generated the following four mutations:

\begin{itemize}
    \setlength{\itemsep}{0pt}
    \item \textbf{Positive---preserve the requirement.}
        A new question asks for the full biographical name of the actor who played Agent J in \emph{Men in Black} and hosted a ceremony featuring Bon Jovi and Toni Braxton.
        The answer is \emph{Willard Carroll Smith II.}; the entity and context change, but the required canonical answer form is preserved.
    \item \textbf{Negative---contrast the requirement.}
        A question about the actress born Caryn Elaine Johnson instead asks for the professional name of the host of the Academy Awards ceremony honoring films released in 1993.
        The answer is \emph{Whoopi Goldberg}, demonstrating a case in which returning the full biographical name would be incorrect.
    \item \textbf{Boundary---change the target.}
        Keeping Michael Andrew Fox and the same Emmy clues, a boundary question explicitly asks for the name by which the actor is known professionally.
        This single change makes \emph{Michael J. Fox} correct and isolates the requested name type as the condition controlling the answer form.
    \item \textbf{Stress---add distractors.}
        A fourth question asks for the exact full name of the musician who wrote ``Manic Monday'' under the pseudonym ``Christopher.''
        The context adds articles about several other people named Prince, but the answer must remain \emph{Prince Rogers Nelson}, testing whether the correction survives misleading name matches.
\end{itemize}

Thus the four operators probe invariance, contrast, decision boundaries, and robustness rather than producing interchangeable paraphrases.
The stored instances retain their full contexts and supporting facts; the questions above are shortened for presentation.

Both the generator and verifier are multi-step agents equipped with task-specific tools.
On HotPotQA, both access \texttt{search\_full\_wiki}, the same full-Wikipedia ColBERT index used by the task program: the generator constructs instances and gold supervision from retrieved passages, while the verifier checks the evidence and can search further.
This grounds synthesis and verification in benchmark evidence rather than the optimizer model's parametric knowledge.

Because tool grounding alone does not guarantee quality, the verifier checks each proposal for \emph{validity} under the task requirements and \emph{faithfulness} to the targeted failure mode and mutation direction.
Only proposals passing both checks enter $D_t$ and provide new evidence for prompt reflection and failure memory.
Additional optimization artifacts, including prompt evolution, candidate lineage, and admitted synthetic instances, are provided in Appendix~\ref{app:artifacts}.

\section{Experiments}
\label{sec:experiments}

Our evaluation addresses three questions.
First, does \forge improve prompt optimization across benchmarks and task models?
Second, is failure-guided synthesis necessary, and how does the synthesis process reshape the failure distribution?
Third, do the resulting synthesized instances benefit optimization methods beyond \forge?

\begin{table*}[!t]
    \centering
    \begingroup
    \fontsize{8pt}{9pt}\selectfont
    \setlength{\tabcolsep}{2pt}
    \renewcommand{\arraystretch}{1.08}
    \begin{tabular*}{\textwidth}{@{\extracolsep{\fill}}lcccccccccc@{}}
        \toprule
        \textbf{Qwen3.5-35B-A3B} & \textbf{HotPotQA} & \textbf{IFBench} & \textbf{HoVer} & \textbf{FiNER-139} & \textbf{LawBench} & \textbf{AIME-2025} & \textbf{MMLU-Pro} & \textbf{AppWorld} & \textbf{Aggregate} & \textbf{Improvement} \\
        \midrule
        Baseline       & 52.00          & 31.29          & 45.00          & 56.00          & 34.00          & 73.33          & 87.00          & 66.64          & 55.66          & --              \\
        \miprov        & 68.00          & 48.64          & 52.00          & 67.00          & 38.00          & 75.56          & \textbf{90.00} & 67.67          & 63.36          & +7.70           \\
        \ace           & 70.00          & \textbf{61.63} & 55.00          & 69.00          & 54.00          & 71.11          & 83.00          & 74.33          & 67.26          & +11.60          \\
        \gepa          & 66.00          & 48.64          & 51.00          & 68.00          & 50.00          & 77.78          & \textbf{90.00} & 71.21          & 65.33          & +9.67           \\
        \forge         & \textbf{76.00} & 54.52          & \textbf{60.00} & \textbf{74.00} & \textbf{57.00} & \textbf{82.22} & 89.00          & \textbf{84.69} & \textbf{72.18} & \textbf{+16.52} \\
        \bottomrule
    \end{tabular*}
    \endgroup
    \caption{Final test performance (\%) with Qwen3.5-35B-A3B. Aggregate averages the eight benchmarks; Improvement is relative to the Baseline. Bold denotes the column best.}
    \label{tab:main-results}
\end{table*}

\subsection{Experimental Setup}
\label{sec:setup}

\paragraph{Benchmarks.}
We evaluate eight heterogeneous LM benchmarks spanning knowledge-intensive reasoning, instruction following, structured prediction, mathematics, and agents.
HotPotQA \cite{yang2018hotpotqa} and HoVer \cite{jiang2020hover} test multi-hop retrieval and evidence aggregation, while IFBench \cite{pyatkin2025ifbench} evaluates generalization to verifiable instructions.
FiNER-139 \cite{loukas2022finer} and LawBench \cite{fei2024lawbench} cover financial entity classification and legal judgment, respectively.
AIME-2025 \cite{matharena2025aime} and MMLU-Pro \cite{wang2024mmlupro} evaluate mathematical and broad knowledge reasoning, and AppWorld \cite{trivedi2024appworld} evaluates interactive agents in stateful tool environments.
Dataset provenance, split construction and task-specific metrics are provided in the appendix~B.

\paragraph{Models and fair comparison.}
Our primary comparison uses Qwen3.5-35B-A3B \cite{qwen2026qwen35} as the task model and GPT-5.5 \cite{openai2026gpt55} as the optimizer model.
We additionally repeat the full comparison with the dense Qwen3-8B task model \cite{yang2025qwen3}.
We compare \forge against the unoptimized baseline and three automatic prompt optimization methods: \miprov \cite{opsahlong2024mipro}, \ace \cite{zhang2026ace}, and \gepa \cite{agrawal2026gepa}.
Within each comparison, all methods use the same task model, training, validation, and test splits, LM-program structure, and task-specific evaluator.
Optimized states are selected using validation performance only, and the test set is accessed only for final evaluation.
To ensure comparable search effort, we align the nominal task-model evaluation budget across optimizers.

\subsection{Main Results}
\label{sec:main-results}

\forge achieves the strongest aggregate performance with Qwen3.5-35B-A3B.
As shown in \Cref{tab:main-results}, it reaches an aggregate score of 72.18, improving over the unoptimized Baseline by 16.52 percentage points and exceeding the competing APO methods by 4.92 to 8.82 points.
\forge obtains the best result on six of eight benchmarks.
The gains therefore span single-module prediction, multi-hop retrieval, mathematical reasoning, and interactive tool use rather than concentrating in one task format.

\subsection{Component Ablation}
\label{sec:component-ablation}

\begin{table}[t]
    \centering
    \begingroup
    \fontsize{8pt}{9pt}\selectfont
    \setlength{\tabcolsep}{2.2pt}
    \renewcommand{\arraystretch}{1.12}
    \begin{tabular*}{\columnwidth}{@{\extracolsep{\fill}}lcccccc@{}}
        \toprule
        \multirow{2}{*}{Variant}
        & Prompt
        & Random
        & Failure
        & Mutation
        & \multirow{2}{*}{Score}
        & \multirow{2}{*}{$\Delta$} \\
        & search
        & synthesis
        & memory
        & guidance
        &
        & \\
        \midrule
        Baseline             &            &            &            &            & 55.66 & --    \\
        + Prompt search      & \checkmark &            &            &            & 64.17 & +8.51 \\
        + Random synthesis   & \checkmark & \checkmark &            &            & 62.38 & -1.79 \\
        + Failure memory     & \checkmark & \checkmark & \checkmark &            & 68.24 & +5.86 \\
        + Mutation guidance  & \checkmark & \checkmark & \checkmark & \checkmark & 72.18 & +3.94 \\
        \bottomrule
    \end{tabular*}
    \endgroup
    \caption{Cumulative ablation averaged over the eight benchmarks in Table~\ref{tab:main-results}.
    Each row adds one component; $\Delta$ is relative to the preceding row.
    Random synthesis generates from randomly selected training instances.}
    \label{tab:ablation-plan}
\end{table}

In \Cref{tab:ablation-plan}, \emph{Score} denotes the mean across the eight benchmarks in \Cref{tab:main-results}.
Prompt search improves the score from 55.66 to 64.17, whereas adding random synthesis lowers it to 62.38.
This contrast shows that merely increasing the training data does not provide the targeted evidence needed for optimization.

Failure guidance makes synthesis effective.
Adding failure memory raises the score to 68.24, surpassing prompt search alone by 4.07 points, and mutation guidance further improves it to 72.18.
Together, these modules outperform random synthesis by 9.80 points and prompt search alone by 8.01 points.
Failure memory identifies \emph{what} weaknesses require new evidence, while mutation guidance determines \emph{how} to expand them into complementary instances.
Because the variants are cumulative, these deltas represent conditional contributions rather than isolated main effects.

\subsection{Synthesis Accounting and Failure Dynamics}
\label{sec:synthesis-accounting}

\begin{figure*}[t]
    \centering
    \includegraphics[width=\textwidth]{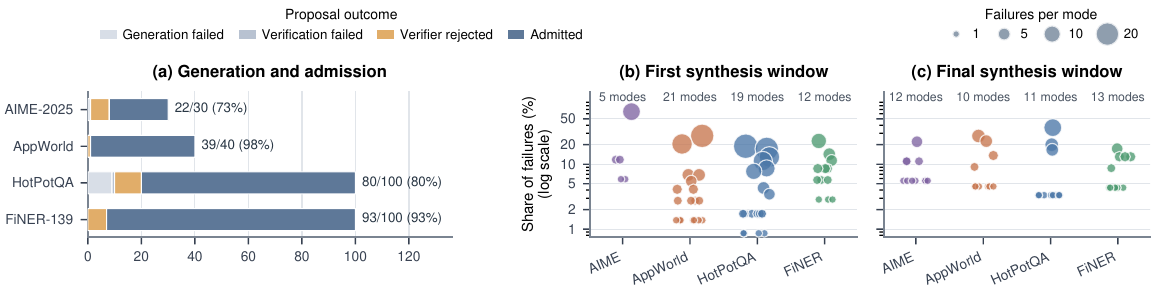}
    \caption{Synthesis accounting and failure-memory dynamics on four representative benchmarks.
    The statistics are computed from the same \forge runs reported in the main comparison (\Cref{tab:main-results}).
    Panel (a) partitions all generation attempts by final outcome; labels report admitted proposals over total attempts and the corresponding admission rate.
    Panels (b) and (c) compare failure modes collected before the first and final synthesis events.
    Each point denotes a task-specific failure mode, vertical position gives its share of failures in the corresponding window on a logarithmic scale, and area gives the number of supporting executions; horizontal jitter has no semantic meaning.
    The comparison highlights substantial redistribution, more comparable support across modes, and benchmark-dependent increases or decreases in mode count.}
    \label{fig:synthesis-accounting}
\end{figure*}

Beyond downstream performance, we audit synthesis outcomes on four representative benchmarks in \Cref{fig:synthesis-accounting}.
Panel (a) accounts for every proposal: \forge admits 22 of 30 on AIME-2025, 39 of 40 on AppWorld, 80 of 100 on HotPotQA, and 93 of 100 on FiNER-139, yielding admission rates of 73--98\%.
Of the ten generation failures, nine arise in HotPotQA because the generator returns no usable output, while the remaining AIME-2025 failure results from an invalid JSON escape.
One additional HotPotQA verification returns non-Boolean validity and faithfulness decisions; these eleven cases reflect malformed outputs rather than quality-based rejection.
Among the 25 successfully verified but rejected proposals, 9 fail validity only, 7 fail faithfulness only, and 9 fail both.
Validity failures primarily involve incorrect supervision or incomplete evidence, whereas faithfulness failures reflect the wrong mutation direction or drift from the targeted failure mode.

Panels (b) and (c) compare failure memory before the first and final synthesis events.
Both mode identities and frequencies change substantially, indicating that later synthesis targets newly exposed weaknesses rather than repeatedly revisiting initial failures.
Within each benchmark, failure counts become more balanced across modes: from the first to the final synthesis window, the ratio between the failure counts of the most and least frequent modes decreases from $11{:}1$ to $4{:}1$ on AIME-2025, $20{:}1$ to $6{:}1$ on AppWorld, $22{:}1$ to $11{:}1$ on HotPotQA, and $8{:}1$ to $4{:}1$ on FiNER-139.
Meanwhile, mode counts increase from 5 to 12 on AIME-2025 and from 12 to 13 on FiNER-139, but decrease from 21 to 10 on AppWorld and from 19 to 11 on HotPotQA.
Thus, the prompt--data loop changes both the composition and balance of failure memory while adapting its granularity to each benchmark.

\subsection{Generalization Across Task Models}
\label{sec:task-model-generalization}

\begin{table*}[!t]
    \centering
    \begingroup
    \fontsize{8pt}{9pt}\selectfont
    \setlength{\tabcolsep}{2pt}
    \renewcommand{\arraystretch}{1.08}
    \begin{tabular*}{\textwidth}{@{\extracolsep{\fill}}lcccccccccc@{}}
        \toprule
        \textbf{Qwen3 8B} & \textbf{HotPotQA} & \textbf{IFBench} & \textbf{HoVer} & \textbf{FiNER-139} & \textbf{LawBench} & \textbf{AIME-2025} & \textbf{MMLU-Pro} & \textbf{AppWorld} & \textbf{Aggregate} & \textbf{Improvement} \\
        \midrule
        Baseline       & 40.00          & 27.34          & 35.00          & 55.00          & 24.00          & 13.33          & 61.00          & 39.34          & 36.88          & --              \\
        \miprov        & 55.00          & 31.75          & 47.00          & 62.00          & 31.00          & 20.00          & 61.00          & 40.64          & 43.55          & +6.67           \\
        \ace           & 61.00          & \textbf{38.39} & 39.00          & 66.00          & \textbf{37.00} & 7.78           & 63.00          & 37.89          & 43.76          & +6.88           \\
        \gepa          & 62.00          & 32.62          & 50.00          & 67.00          & 30.00          & 18.89          & 61.00          & 43.23          & 45.59          & +8.72           \\
        \forge         & \textbf{70.00} & 35.81          & \textbf{57.00} & \textbf{72.00} & 35.00          & \textbf{23.33} & \textbf{66.00} & \textbf{48.64} & \textbf{50.97} & \textbf{+14.10} \\
        \bottomrule
    \end{tabular*}
    \endgroup
    \caption{Final test performance (\%) with dense Qwen3-8B.
    Aggregate and Improvement follow Table~\ref{tab:main-results}; bold denotes the column best.}
    \label{tab:task-model-ablation}
\end{table*}

The gains of \forge persist when the task model changes from the mixture-of-experts Qwen3.5-35B-A3B to the dense Qwen3-8B.
On Qwen3-8B, \forge leads six of eight benchmarks and achieves an aggregate score of 50.97, exceeding the Baseline by 14.10 points and the strongest competing optimizer, \gepa, by 5.38 points (\Cref{tab:task-model-ablation}).
Together with its 16.52-point gain over the Baseline and 4.92-point lead over the strongest competing APO method on Qwen3.5-35B-A3B, these results suggest that the advantage of \forge is consistent across dense and mixture-of-experts architectures.
We further test the robustness of \forge to optimizer-model choice in Appendix~\ref{app:supplementary-experiments}.

\subsection{Transferability of Synthesized Data}
\label{sec:data-transfer}

We next test whether \forge-generated data remain useful after changing the optimizer that consumes them.
On HotPotQA, FiNER-139, and LawBench, we compare the original training data with the same data augmented by \forge-synthesized instances for three prompt optimizers and for GRPO \cite{shao2024deepseekmath}.
All comparisons use Qwen3.5-35B-A3B as the task model, and the three prompt optimizers use GPT-5.5 as the optimizer model.
Each method keeps its optimization budget fixed across the two data conditions.
The two GRPO arms differ only in their training data, using the same number of optimization steps and otherwise identical configurations.
Appendix~\ref{app:grpo-setup} provides the complete GRPO protocol and configuration.
Augmentation improves all twelve benchmark--optimizer pairs, showing that the synthesized instances remain useful beyond \forge.

\begin{table}[!htbp]
    \centering
    \begingroup
    \footnotesize
    \setlength{\tabcolsep}{1.5pt}
    \renewcommand{\arraystretch}{1.12}
    \begin{tabular}{@{}llcccc@{}}
        \toprule
        & & \multicolumn{3}{c}{Prompt Optimization} & \multicolumn{1}{c}{\shortstack{Weight \\ Optimization}} \\
        \cmidrule(lr){3-5}\cmidrule(l){6-6}
        Dataset & Training data & \miprov & \ace & \gepa & GRPO \\
        \midrule
        \multirow{2}{*}{HotPotQA}
            & Original & 68 & 70 & 66 & 57 \\
            & $+\,$\forge data & \textbf{73} & \textbf{73} & \textbf{75} & \textbf{65} \\
        \midrule
        \multirow{2}{*}{FiNER-139}
            & Original & 67 & 69 & 68 & 69 \\
            & $+\,$\forge data & \textbf{70} & \textbf{71} & \textbf{74} & \textbf{73} \\
        \midrule
        \multirow{2}{*}{LawBench}
            & Original & 38 & 54 & 50 & 43 \\
            & $+\,$\forge data & \textbf{47} & \textbf{57} & \textbf{55} & \textbf{51} \\
        \bottomrule
    \end{tabular}
    \endgroup
    \caption{Test performance (\%) with original versus \forge-augmented training data under matched configurations and budgets.
    Bold denotes the better condition.}
    \label{tab:data-transfer-comparison}
\end{table}

\Cref{tab:data-transfer-comparison} shows that \forge augmentation improves all twelve dataset--optimizer pairs: the nine prompt-optimization gains span 2--9 points, while GRPO gains at step 100 are 8, 4, and 8 points on HotPotQA, FiNER-139, and LawBench, respectively.
These results show that the synthesized instances transfer beyond \forge to other prompt optimizers and model-weight optimization.

\begin{figure}[t]
    \centering
    \includegraphics[width=\columnwidth]{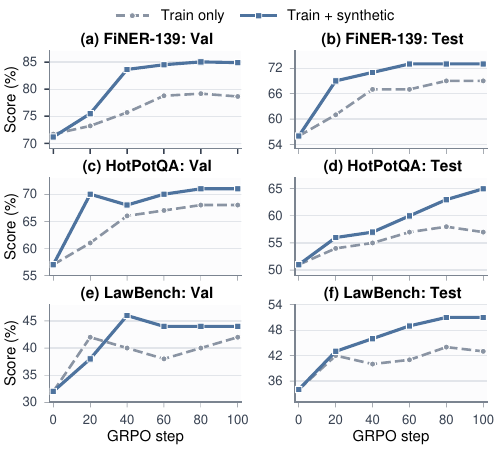}
    \caption{GRPO validation and test trajectories on FiNER-139, HotPotQA, and LawBench.
    The two arms differ only in whether the training data are augmented with \forge-synthesized instances.}
    \label{fig:grpo-training-dynamics}
\end{figure}

\Cref{fig:grpo-training-dynamics} separates validation and test trajectories for all three GRPO comparisons.
Under the same 100-step optimization budget, the augmented arm finishes above original-data training on all three test sets.
The corresponding validation trajectories show that these gains emerge at different stages and need not increase monotonically.
Together, the six panels show that \forge-synthesized instances provide reusable training evidence across three distinct weight-optimization tasks.

\section{Conclusion}
\label{sec:conclusion}

\forge uses failures as shared signals for reflective prompt updates and targeted data synthesis through failure memory, four mutation intents, and agentic verification.
Across eight benchmarks, it outperforms the unoptimized baseline and all evaluated APO methods; ablations show that these gains require failure- and mutation-guided rather than random synthesis.
Its instances also improve three competing prompt optimizers and GRPO across three transfer benchmarks, demonstrating that the synthesized data remain useful beyond the optimization loop that produced them.
Together, these results suggest that prompt search and data construction should be treated as a coupled process: observed failures reveal both how a prompt should change and which training examples should be added next.
Future work should broaden this transfer evaluation across additional tasks and model families and disentangle the effects of dataset size, example selection, and sampling distribution.

\clearpage
\bibliography{main}

\clearpage
\appendix
\setcounter{secnumdepth}{2}
\renewcommand{\thesection}{\Alph{section}}
\renewcommand{\thesubsection}{\thesection.\arabic{subsection}}
\crefname{section}{Appendix}{Appendices}
\Crefname{section}{Appendix}{Appendices}
\crefname{subsection}{Appendix}{Appendices}
\Crefname{subsection}{Appendix}{Appendices}

\makeatletter
\def\addcontentsline#1#2#3{%
    \addtocontents{#1}{\protect\contentsline{#2}{#3}{\thepage}{}%
    \protected@file@percent}}
\makeatother

\setcounter{tocdepth}{2}
\renewcommand{\contentsname}{Appendix Contents}
\tableofcontents
\clearpage

\section{Algorithm and Implementation Details}
\label{app:algorithm}

\Cref{alg:forge} in the main paper gives the complete prompt--data optimization loop but leaves its internal calls abstract.
This section instantiates exactly the six operations used in that loop: optimization state, parent-candidate selection, minibatch sampling, failure-memory construction, reflective prompt update, and failure-guided synthesis with verification.
We describe decision-level behavior and the model-facing contracts needed for reproduction, while omitting checkpoint and callback bookkeeping.

\subsection{Optimization State}
\label{app:optimization-state}

The conceptual optimization state is $(\Pi,D,S_V,\mathcal M)$.
The candidate pool $\Pi=[\boldsymbol{\pi}_0,\ldots,\boldsymbol{\pi}_{q-1}]$ stores complete mappings from program-module names to system prompts; an accepted candidate is appended to this pool rather than replacing its parent.
The mutable training data $D$ contain the original instances and all admitted synthetic instances.
For every candidate $i$, $S_V$ stores its per-instance validation scores $s_i(z)$ for all $z\in V$.
These scores support both parent selection during search and aggregate validation-based model selection.

Failure memory $\mathcal M$ groups imperfect executions into reusable failure modes.
A mode is $m_j=(n_j,d_j,O_j)$, where $n_j$ is a short name, $d_j$ is a decontextualized mechanism description, and $O_j$ is the ordered collection of observations supporting that description.
Each observation retains the parent-candidate index, task instance, scalar score, ordered program trace, and module-level feedback.
The implementation stores modes by synthesis epoch: earlier epochs remain available for audit, but only modes collected since the preceding synthesis event are active when constructing the next synthetic batch.

\subsection{Candidate Selection}
\label{app:candidate-selection}

\forge uses validation behavior to select a parent without collapsing every candidate to one aggregate score.
For each validation instance $z$, it forms the local front
\begin{equation}
    F_z=\left\{i:
    s_i(z)=\max_{\ell}s_{\ell}(z)\right\}.
\end{equation}
A candidate $i$ is placed in the dominated set $G$ when some candidate
$\ell\neq i$ satisfies $s_{\ell}(z)\geq s_i(z)$ for every $z\in V$ and is
strictly better on at least one validation instance.
The default selector then samples uniformly from the multiset of non-dominated front members, as detailed in \Cref{alg:select-candidate}.

\begin{algorithm}[!t]
\caption{Front-frequency candidate selection}
\label{alg:select-candidate}
\begin{algorithmic}[1]
\REQUIRE candidate pool $\Pi$; per-instance validation scores $S_V$; validation set $V$
\STATE $G\gets\{i:\Call{Dominated}{i,S_V}\}$
\STATE $L\gets[\,]$ \COMMENT{multiset of eligible front members}
\FOR{each validation instance $z\in V$}
    \STATE $b_z\gets\max_{\ell}s_{\ell}(z)$
    \FOR{each candidate index $i=0,\ldots,|\Pi|-1$}
        \IF{$i\notin G$ and $s_i(z)=b_z$}
            \STATE append $i$ to $L$
        \ENDIF
    \ENDFOR
\ENDFOR
\STATE $i^\star\gets\Call{UniformChoice}{L}$
\RETURN parent candidate $\Pi[i^\star]$
\end{algorithmic}
\end{algorithm}

Equivalently, candidate $i$ has multiplicity
\begin{equation}
    w_i=\mathbf{1}[i\notin G]\sum_{z\in V}\mathbf{1}[i\in F_z],
    \qquad
    P(i)=\frac{w_i}{\sum_{\ell}w_{\ell}}.
\end{equation}
Thus a candidate that leads on more validation instances is sampled more often, while complete domination removes it from consideration.

\subsection{Minibatch Sampling}
\label{app:minibatch-sampling}

The default sampler is epoch-shuffled.
It stores a shuffled order of training-instance identifiers and a cursor in that order, then returns the next contiguous minibatch at each prompt-search step.
When the order is exhausted, all identifiers are reshuffled for a new sampling epoch.
If $|D|$ is not divisible by the minibatch size, the final order is padded with randomly selected identifiers so that every returned minibatch is full; only these padding positions may duplicate an instance.

A synthesis event refreshes the order without discarding which instances have already been consumed.
The sampler first collects unique, not-yet-consumed identifiers from the current order and adds identifiers of newly admitted synthetic instances.
It shuffles this group and places it before a separately shuffled group of already-consumed identifiers.
Consequently, new data receive the same priority as unseen real data and enter subsequent prompt updates before instances already visited in the current sampling epoch.
This sampling epoch is independent of the synthesis epoch used to partition failure memory.

\subsection{Failure-Memory Construction}
\label{app:failure-memory-construction}

\begin{figure*}[!t]
    \centering
    \includegraphics[width=0.95\textwidth]{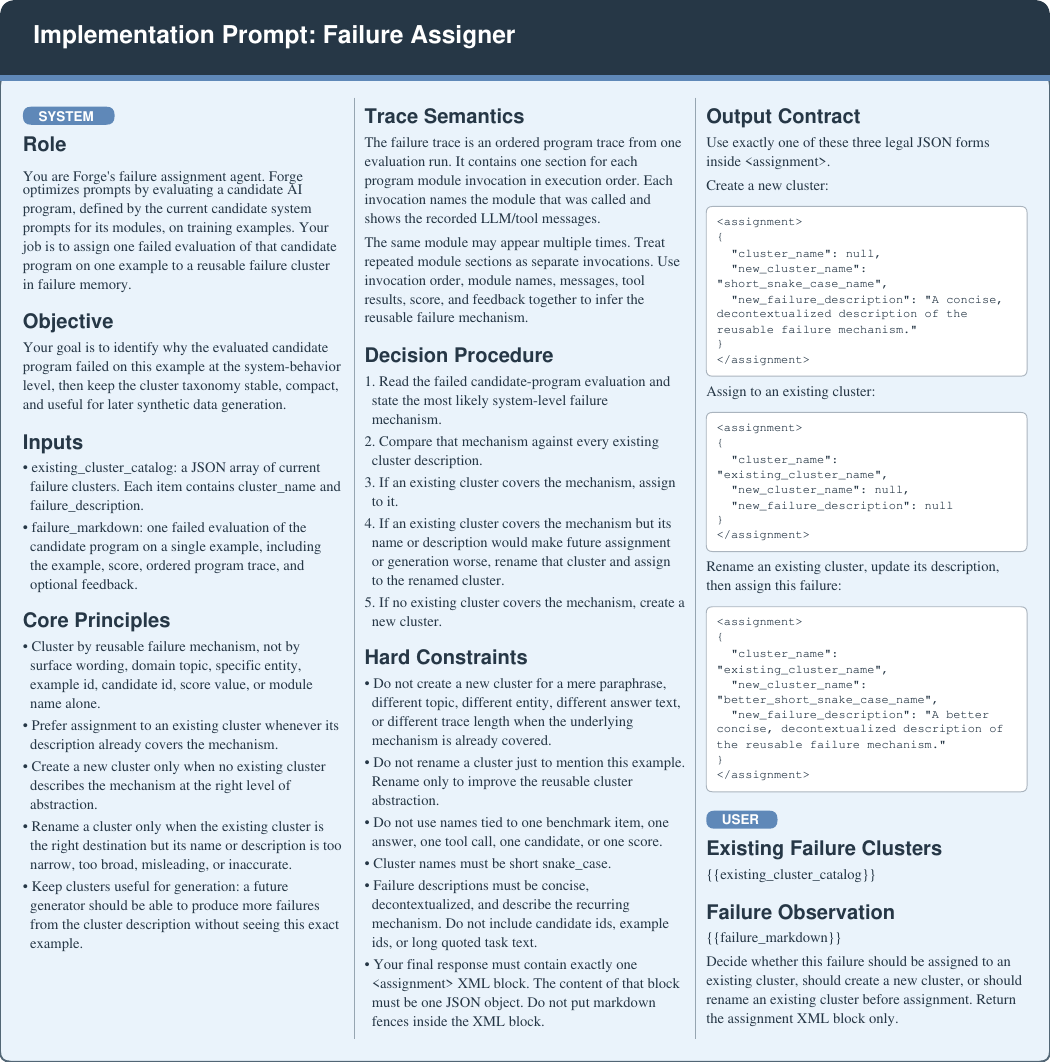}
    \caption{The failure-assigner prompt maps a failed execution to an existing or newly created failure mode.}
    \label{fig:failure-assigner-prompt}
\end{figure*}

Failure memory is built only from imperfect executions of the selected parent on the current training minibatch.
For each evaluation with score below $1$, \forge constructs an observation
$o=(i,z,r,\tau,f)$ containing the parent index $i$, instance $z$, score $r$, complete ordered program trace $\tau$, and module-level feedback $f$.
The assigner receives two textual inputs: a catalogue containing only the names and descriptions of modes in the current synthesis epoch, and a Markdown rendering of the new observation with sections for the instance, score, trace, and optional feedback.
\Cref{fig:failure-assigner-prompt} shows the complete assigner prompt and its dynamic user fields.
The implementation calls a failure mode a \emph{failure cluster}; we retain the paper terminology in the prose.

The assigner returns one of three structured actions.
\emph{Assign} appends the observation to an existing mode; \emph{create} introduces a new name and description with the observation as its first support; and \emph{rename} updates an existing name and description before appending the observation.
The rename action is the implementation of the abstraction refinement described in the main paper and may also keep the name unchanged while improving only the description.
Failures are processed sequentially, so a later failure in the same minibatch sees every mode created or revised by earlier failures.
Malformed JSON, an illegal action-field combination, an unknown destination, or a naming conflict produces a no-op rather than an ungrounded memory update.

The resulting catalogue provides stable synthesis targets, while each support set retains the executions that justify its target.
After synthesis, \forge advances the synthesis epoch.
The stored history is preserved, but the next active catalogue starts empty so that future generation reflects weaknesses exposed by the updated prompt--data state.

\subsection{Reflective Prompt Update}
\label{app:prompt-update}

\begin{figure*}[!t]
    \centering
    \includegraphics[width=0.95\textwidth]{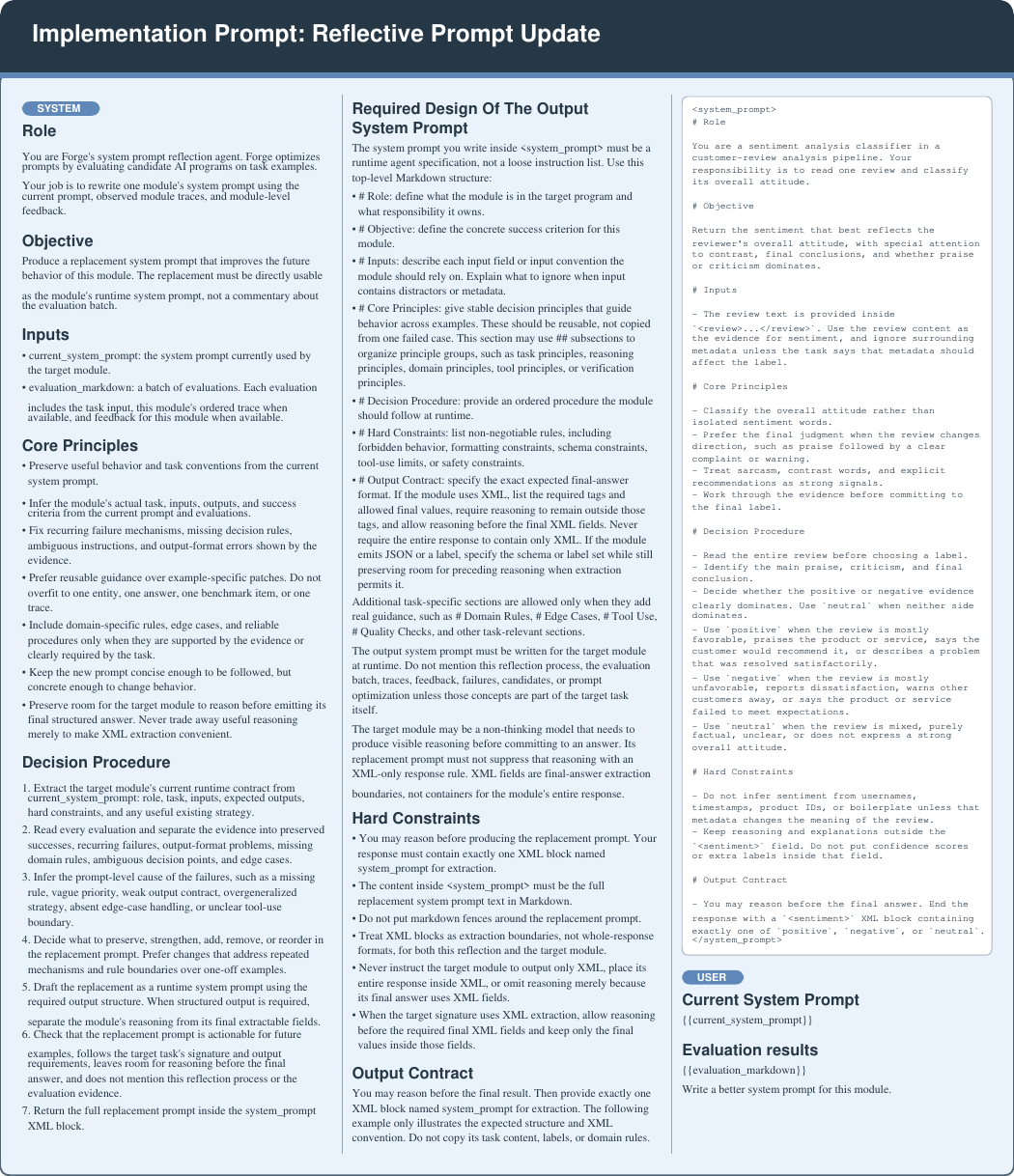}
    \caption{The reflective prompt-update prompt produces a complete replacement system prompt from imperfect module executions.}
    \label{fig:reflection-prompt}
\end{figure*}

The prompt-update branch uses the imperfect parent executions already collected for the current minibatch.
Under the default round-robin policy, step $t$ selects one program module, and the reflector receives that module's current system prompt together with a Markdown bundle containing each failed task input, the trace entries belonging to that module, and its module-level feedback.
The reflector proposes one complete replacement system prompt.
If this block is absent or empty, the module retains its original prompt.
\Cref{fig:reflection-prompt} shows this model-facing reflection contract.

The proposal copies all unselected module prompts from the parent and replaces only the selected module.
Although reflection uses only imperfect evaluations, parent and proposal are screened on the same \emph{complete} minibatch.
The proposal is appended to $\Pi$ only when its total minibatch score is strictly higher; an admitted proposal is then evaluated on all of $V$ and added to $S_V$.
This local paired comparison prevents an edit from being accepted merely because it was tested on an easier batch.

\subsection{Failure-Guided Synthesis and Verification}
\label{app:synthesis-details}

\begin{figure*}[!t]
    \centering
    \includegraphics[width=0.95\textwidth]{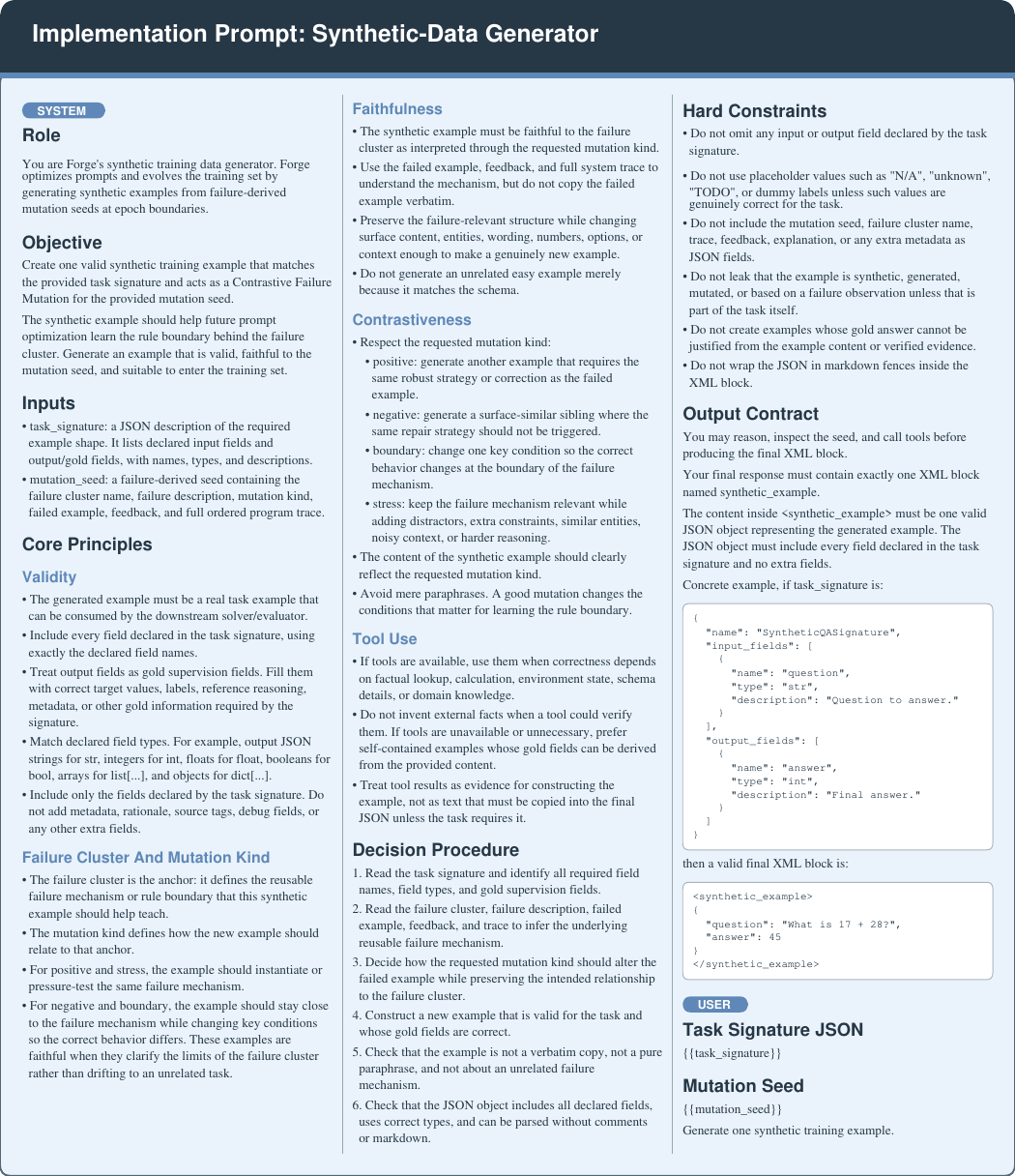}
    \caption{The synthetic-data generator prompt produces a failure-guided training example from a task signature and mutation seed.}
    \label{fig:generator-prompt}
\end{figure*}

\begin{figure*}[!t]
    \centering
    \includegraphics[width=0.95\textwidth]{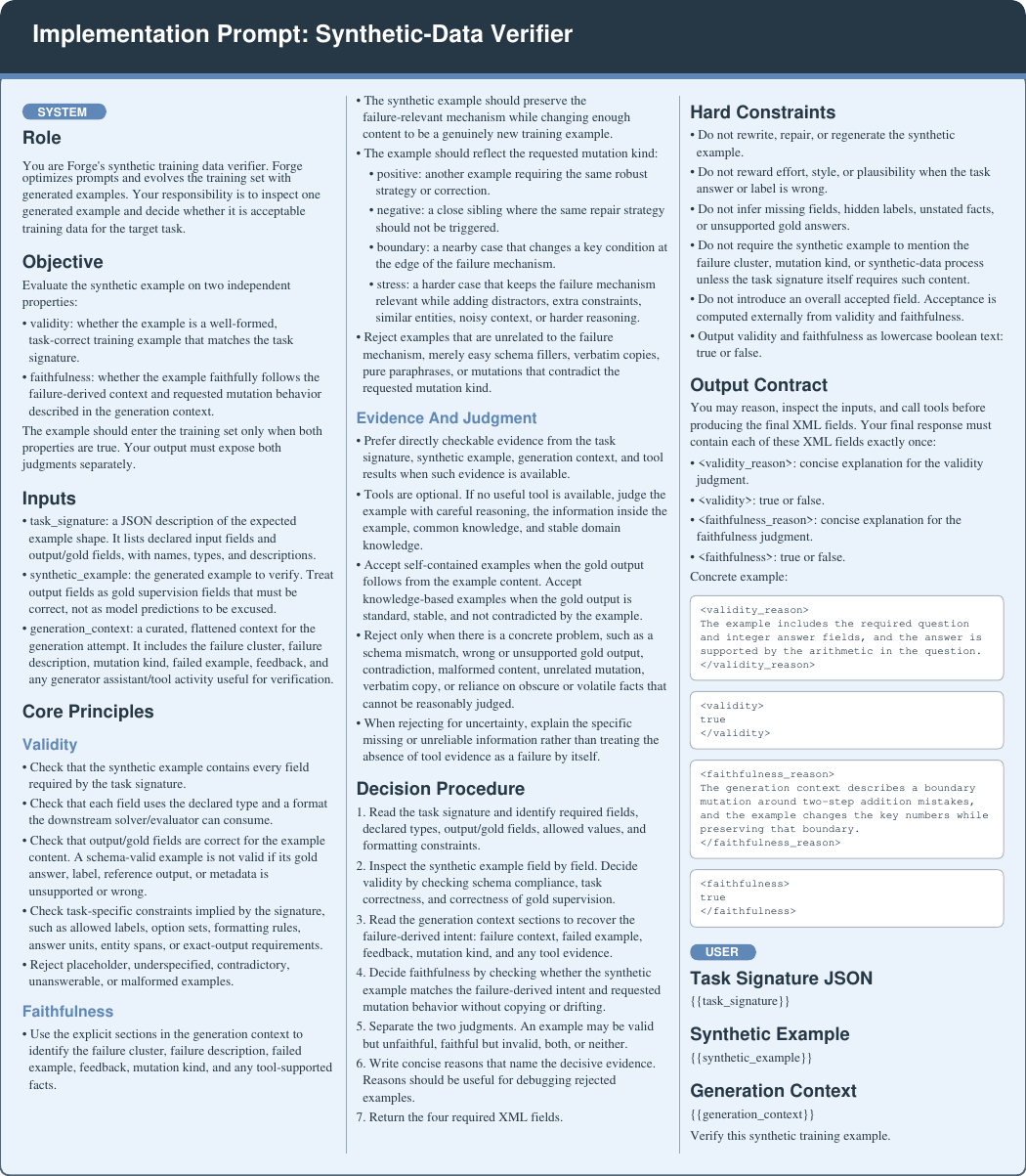}
    \caption{The synthetic-data verifier prompt independently judges the validity and failure-mode faithfulness of a generated example.}
    \label{fig:verifier-prompt}
\end{figure*}

The exact model-facing generator and verifier contracts are shown in
\Cref{fig:generator-prompt,fig:verifier-prompt}; the admission logic connecting them is specified below.

A synthesis event is triggered after $p$ completed prompt-search steps without a strict improvement in the historical best aggregate validation score.
The event budget $K$ counts \emph{generation attempts}, not successfully parsed or finally admitted instances.
Let $N=\sum_j|O_j|$ be the number of observations supporting the active modes.
\forge first assigns
\begin{equation}
    k_j^{(0)}=\left\lfloor K\frac{|O_j|}{N}\right\rfloor
\end{equation}
attempts to mode $j$, then distributes the remaining attempts in descending order of the fractional remainders, breaking equal remainders by mode order.
This largest-remainder allocation preserves $\sum_j k_j=K$ while prioritizing recurrent failures.

Within each allocated slot, \forge samples one supporting observation uniformly with replacement.
It assigns mutation intents from the global cycle
$\Omega=(\textsc{Positive},\textsc{Negative},\textsc{Boundary},\textsc{Stress})$ across the complete seed list, restarting from \textsc{Positive} at each synthesis event.
\Cref{alg:failure-guided-synthesis} gives the complete admission path.

\begin{algorithm}[!t]
\caption{Failure-guided synthesis, verification, and admission}
\label{alg:failure-guided-synthesis}
\begin{algorithmic}[1]
\REQUIRE active modes $\mathcal M=\{(n_j,d_j,O_j)\}$; task signature $\Sigma$; attempt budget $K$; training data $D$
\STATE $\Omega\gets(\textsc{Positive},\textsc{Negative},\textsc{Boundary},\textsc{Stress})$
\STATE $(k_1,\ldots,k_{|\mathcal M|})\gets\Call{LargestRemainder}{K,(|O_j|)_j}$
\STATE $A\gets\varnothing$; $c\gets0$
\FOR{each mode $m_j=(n_j,d_j,O_j)\in\mathcal M$}
    \FOR{$a=1,\ldots,k_j$}
        \STATE $o\gets\Call{UniformChoice}{O_j}$ \COMMENT{sampling with replacement}
        \STATE $u\gets\Omega[1+(c\bmod|\Omega|)]$; $c\gets c+1$
        \STATE $(\widetilde z,\tau_G)\gets\Call{Generate}{\Sigma,m_j,o,u}$
        \IF{$\Call{Coerce}{\widetilde z,\Sigma}=\bot$}
            \STATE \textbf{continue} \COMMENT{skip verifier on parse/schema failure}
        \ENDIF
        \STATE $C\gets(m_j,u,o.\mathrm{instance},o.\mathrm{feedback})$
        \STATE append generator assistant/tool activity from $\tau_G$ to $C$
        \STATE $(v,f)\gets\Call{Verify}{\Sigma,\widetilde z,C}$
        \IF{$v=\mathrm{true}$ and $f=\mathrm{true}$}
            \STATE $A\gets A\uplus\{\widetilde z\}$
        \ENDIF
    \ENDFOR
\ENDFOR
\STATE $D\gets D\uplus A$; \Call{RefreshSampleOrder}{$D$}; advance synthesis epoch
\RETURN admitted synthetic data $A$
\end{algorithmic}
\end{algorithm}

The generator is a multi-step agent conditioned on the JSON task signature and a mutation seed containing the mode name and description, mutation intent, failed instance, feedback, and full source program trace.
It may call benchmark-provided tools when factual lookup, calculation, environment state, or schema details require grounding.
Its final XML field must contain a JSON object.
The implementation parses that object and coerces all declared fields and types before verification; a failure at this stage is rejected immediately.

The verifier is a second multi-step agent equipped with the same benchmark generation tools.
It receives the task signature, parsed synthetic instance, and a flattened context containing the mode, intent, failed instance, source feedback, and the generator's assistant/tool activity.
It does not directly receive the source program trace or hidden generator reasoning.
The verifier independently judges \emph{validity}---schema compliance, task coherence, and gold-supervision correctness---and \emph{faithfulness} to both the failure mode and mutation intent.
Only two successfully parsed Boolean values that are both true admit an instance.

\section{Experimental Setup}
\label{app:experimental-setup}

This section specifies the data, evaluators, models, prompt-optimization budget alignment, and the separate paired GRPO transfer protocol used in the main paper.

\subsection{Benchmark Data and Splits}
\label{app:experimental-data}

The comparison protocol assigns the same ordered training, validation, and test examples to every optimizer for a given benchmark and task model.
Unless a benchmark defines a fixed split, randomized construction uses the stated local seed and produces disjoint partitions.
\Cref{tab:experimental-splits} summarizes the exact construction and number of examples in each benchmark split.

\begin{table*}[t]
    \centering
    \caption{Dataset construction used in the eight-benchmark comparisons.}
    \label{tab:experimental-splits}
    \small
    \setlength{\tabcolsep}{3.2pt}
    \renewcommand{\arraystretch}{1.12}
    \begin{tabular}{lrrrp{9.4cm}}
        \toprule
        Benchmark & Train & Val & Test & Source and split construction \\
        \midrule
        AIME
            & 45 & 45 & 90
            & Shuffle the 90 AIME 2022--2024 problems from the AI-MO AIMO Validation AIME dataset with seed 0 and split them equally into train and validation \cite{aimo2024validationaime}; use all 30 problems from the MathArena AIME-2025 dataset for test and evaluate each three times \cite{matharena2025aime}. \\
        AppWorld
            & 90 & 57 & 168
            & Use the complete AppWorld \texttt{train}, \texttt{dev}, and \texttt{test\_normal} task lists, respectively, in their stored order and without difficulty filtering \cite{trivedi2024appworld}. \\
        FiNER-139
            & 200 & 100 & 100
            & Use the fixed \texttt{train}/\texttt{val}/\texttt{test} files distributed with the Meta-Harness FiNER adapter, preserving file order \cite{loukas2022finer,lee2026metaharness}. \\
        HotPotQA
            & 100 & 100 & 100
            & Partition the official \texttt{fullwiki} training split by position into the first 40\% (test pool), middle 40\% (validation pool), and final 20\% (training pool), then sample 100 examples from each pool with seed 0 \cite{yang2018hotpotqa}. \\
        HoVer
            & 100 & 100 & 100
            & Draw train from the official training data with seed 0; construct mutually exclusive validation and test sets from the official development data with seed 1.  Each split contains a near-uniform 34/33/33 mixture of 2-, 3-, and 4-hop claims \cite{jiang2020hover}. \\
        IFBench
            & 150 & 150 & 294
            & Divide the 14,971-row bundled training file into 150 contiguous segments; with seed 1, sample two distinct rows per segment and assign one to train and one to validation.  Use all rows in the fixed test file \cite{pyatkin2025ifbench}. \\
        LawBench
            & 200 & 50 & 100
            & Use the fixed \texttt{train}/\texttt{val}/\texttt{test} crime-prediction files distributed by the MCE artifact, preserving file order \cite{fei2024lawbench,ye2026mce}. \\
        MMLU-Pro
            & 100 & 70 & 100
            & From the official test split, sample mutually exclusive train and test sets with seed 0 and category-proportional largest-remainder quotas; use the complete official validation split \cite{wang2024mmlupro}. \\
        \bottomrule
    \end{tabular}
\end{table*}

\subsection{Benchmark Evaluation, Tools, and Synthesis}
\label{app:experimental-evaluation}

Under this protocol, every comparison method calls the same benchmark adapter and receives the same scalar score for a given program execution.
Validation scores select optimized states, whereas test outcomes are retained only for reporting and never provide feedback to optimization.
For each benchmark, the reported percentage is 100 times the arithmetic mean of its per-example test scores; the aggregate in the main tables is the unweighted mean of the eight benchmark percentages.
We report adapter-provided resources alongside each evaluator because they constrain optimization-side generation and verification; unless stated otherwise, they do not change the scalar metric.
Credentials and machine-specific configuration are not embedded in any prompt.

\noindent\textbf{Synthesis schedule.}
Following \cref{app:synthesis-details}, we denote each benchmark's configuration by $(K,E,P)$, where $K$ is the number of generation attempts requested at each synthesis event, $E$ is the maximum number of synthesis epochs, and $P$ is the number of completed prompt-search steps without a strict improvement in the historical best aggregate validation score that triggers the next event.
The configured tuples are AIME $(10,3,15)$; AppWorld $(10,4,15)$; FiNER-139 $(25,4,15)$; HotPotQA $(25,4,15)$; HoVer $(25,4,15)$; IFBench $(25,4,15)$; LawBench $(25,4,15)$; and MMLU-Pro $(25,4,15)$.
Thus, the configured attempt ceilings are 30 for AIME, 40 for AppWorld, and 100 for each remaining benchmark.
The value $K$ counts mutation-seed generation attempts, not admitted samples; parsing and verification failures can reduce admissions.

\noindent\textbf{AIME.}
The evaluator strips whitespace, parses the prediction as an integer, and applies exact match.
The adapter exposes no benchmark-specific external lookup tool; reference solutions are used only to construct feedback.
\par\noindent
\textbf{AppWorld.}
The score is the fraction of benchmark state-based tests passed after interaction with the persistent environment, with at most 50 agent steps; an execution error receives zero.
For optimization-side synthesis, the adapter exposes an authoring service that can inspect APIs, draft and repair native tasks, validate solutions, and finalize published task artifacts.
The generic optimizer passes this authoring interface to both generator and verifier; although a separate read-only inspection interface exists, the present runs do not implement verification as an independently read-only path.

\noindent\textbf{FiNER-139.}
The predicted tag is compared case-insensitively with the gold tag over the adapter's fixed 12-label XBRL vocabulary.
The optimization-side agents also receive the adapter's label-concept resource.
\par\noindent
\textbf{HotPotQA.}
The primary score is normalized short-answer exact match after two retrieval hops.
During synthesis, \texttt{search\_full\_wiki} queries the same full-Wikipedia ColBERT index used by the task program.
The generator uses retrieved passages to construct the question, answer, aligned context, and supporting-fact indices; the verifier receives the complete retrieval trace and may issue additional queries when factual support requires checking.

\noindent\textbf{HoVer.}
The binary score is one iff all normalized gold supporting-document titles occur among three top-seven retrievals.
Extra documents are not penalized, and the claim label is not predicted.
The adapter exposes the corresponding full-wiki search interface for claims and evidence documents.
\par\noindent
\textbf{IFBench.}
The score is the fraction of executable constraints satisfied; each checker accepts any of eight formatting-normalized variants obtained by optionally removing asterisks and boundary lines.
The optimization-side agents receive constraint metadata together with the official instruction rendering.

\noindent\textbf{LawBench.}
The evaluator applies unordered exact match to the predicted and gold sets of semicolon-separated criminal charges.
The adapter also supplies the official charge catalog.
\par\noindent
\textbf{MMLU-Pro.}
Accuracy is computed after extracting one standalone answer letter from A through J.
The adapter exposes no benchmark-specific external lookup tool.

\subsection{Model Configuration}
\label{app:experimental-models}

The experiments separate the frozen \emph{task model}, which executes the benchmark program, from the \emph{optimizer model}, which proposes optimization-side changes.

\noindent\textbf{Primary task model.}
The eight-benchmark comparison uses Qwen3.5-35B-A3B \cite{qwen2026qwen35}, with a maximum completion length of 8,192 tokens and thinking disabled.
The cross-architecture comparison replaces it with Qwen3-8B \cite{yang2025qwen3}, using the same completion limit and non-thinking setting.

\noindent\textbf{Optimizer model.}
GPT-5.5 is used with medium reasoning effort.
\forge uses it for reflection, failure assignment, synthetic-example generation, and verification; GEPA uses it for reflection, MIPROv2 for proposal, and ACE for reflection and curation.

All methods keep the task-model weights fixed and share the stated Qwen inference settings, LM-program structure, and evaluator.
All other decoding parameters, including temperature, top-$p$, and top-$k$, follow the official Qwen defaults.

\subsection{Optimization-Budget Alignment}
\label{app:experimental-budget}

The configured comparison uses seed 0, the complete training split and validation-only state selection.

Because the optimizers expose different iteration units, we use completed per-example task-program evaluations as the common search-effort proxy.
For each static benchmark and task model, we first run \forge to completion and use its completed count as the reference ceiling.
We pass that ceiling to GEPA and ACE when their APIs permit; for MIPROv2, we choose the largest conservative trial count whose estimated task-program evaluations do not exceed the ceiling.
The unoptimized Baseline performs no search evaluations and is therefore a control rather than a cost-matched optimizer.

This protocol aligns a common task-program evaluation ceiling rather than exact realized compute: GEPA may overshoot at iteration boundaries, MIPROv2's estimate excludes proposal-stage calls, and ACE may stop early under its native rule.

\subsection{GRPO Transfer Protocol}
\label{app:grpo-setup}

The weight-optimization study asks whether examples synthesized by \forge remain useful when the learner updates model weights rather than prompts.
It is a paired data intervention, not a component of the \forge optimizer, and uses group relative policy optimization (GRPO) \cite{shao2024deepseekmath}.

\noindent\textbf{Paired data intervention.}
For every reported benchmark, both arms start from the same Qwen3.5-35B-A3B checkpoint \cite{qwen2026qwen35} and run serially with seed and data seed 42.
The \texttt{train\_only} arm uses the original training split, whereas \texttt{train\_plus\_synthetic} appends a fixed set of verifier-admitted \forge examples.
The two arms use identical validation and test files, and no resampling or explicit real/synthetic mixture is applied.
The real/synthetic/augmented training counts are 200/93/293 for FiNER-139, 100/80/180 for HotPotQA, and 200/88/288 for LawBench; their validation/test counts are 768/100, 100/100, and 50/100, respectively.

\noindent\textbf{Shared optimization configuration.}
Both arms run for exactly 100 optimizer steps, rather than a fixed number of epochs, and differ only in training-set composition.
Each arm therefore consumes 6,400 completion-level training instances, organized as 800 groups of eight completions.
The principal hyperparameters shared by both arms are summarized in \cref{tab:grpo-config}.

\begin{table}[!t]
    \centering
    \caption{GRPO configuration shared by the original-data and augmented-data arms.}
    \label{tab:grpo-config}
    \small
    \begin{tabular}{lr}
        \toprule
        Setting & Value \\
        \midrule
        Random / data seed & 42 / 42 \\
        Train steps & 100 \\
        Global / micro batch & 64 / 1 \\
        Generations per prompt & 8 \\
        Steps per generation (single / agent) & 2 / 1 \\
        Learning rate & $5\times10^{-5}$ \\
        Schedule / warmup & cosine / 0 \\
        GRPO $\beta$ & 0 \\
        Clip low / high & 0.20 / 0.28 \\
        Temperature / top-$p$ / top-$k$ & 1.0 / 1.0 / $-1$ \\
        Completion / context limit & 8,192 / 64,000 \\
        LoRA rank / alpha / dropout & 8 / 32 / 0.05 \\
        \bottomrule
    \end{tabular}
\end{table}

\noindent\textbf{Experimental platform.}
The GRPO runs use Ubuntu 20.04 with eight 48-GB NVIDIA RTX A6000 GPUs.
The software stack comprises Python 3.12.4, \texttt{forge-grpo} 0.1.0, PyTorch 2.10.0 with CUDA 12.8.
Training uses BF16; colocated vLLM rollout uses tensor parallelism 4 and a GPU-memory-utilization target of 0.6.
Use Adam with $\beta=(0.9,0.95)$, $\epsilon=10^{-8}$, weight decay 0.1, and gradient clipping at 1.0.
Optimizer and RNG state are retained for strict resume, and LoRA adapters are not merged into the base model.

\noindent\textbf{Task-specific rollout and reward.}
FiNER-139 and LawBench use single-turn rollouts.
Their reward is the benchmark score defined in \cref{app:experimental-evaluation} with weight 1.0 plus a soft XML-format reward with weight 0.1; because the benchmark parsers also accept bare outputs, XML is encouraged rather than required for task correctness.
Their 128-completion generation batch contains 16 prompt groups and supplies two consecutive optimizer steps.

HotPotQA instead runs a continuous search-agent trajectory and regenerates a 64-completion batch at every optimizer step.
Each of its eight prompt groups contains eight trajectories, each trajectory permits at most three actions, and every model turn has an independent 8,192-token completion allowance.
Each search retains at most seven passages, truncated to 700 characters each.
The trajectory reward is
\begin{equation}
    \begin{aligned}
        r_{\mathrm{HotPotQA}}
        ={}&0.8\,\mathrm{EM}_{\mathrm{terminal}} \\
        &+0.2\sum_t\max\!\left(0,\mathrm{Recall}_t-\mathrm{Recall}_{t-1}\right),
    \end{aligned}
\end{equation}
where EM is normalized answer exact match and recall is supporting-title recall.
No separate static reward is added, and evaluation reports the benchmark's primary answer EM rather than this shaped training reward.
The retriever corpus, index, and top-seven interface are fixed across arms and checkpoints.
The reported comparison uses the step-100 test endpoint; complete held-out splits are evaluated, test outcomes are monitoring-only, and training always reaches 100 updates.

\section{Optimization Artifacts}
\label{app:artifacts}

\makeatletter
\setlength{\@dblfptop}{0pt}
\makeatother

This section documents three optimization artifact types: an AppWorld prompt before and after optimization, a HoVer candidate lineage, and admitted synthetic examples.
The figures are generated directly from the corresponding checkpoint and result files; they are illustrations of stored state rather than additional evaluations.

\subsection{AppWorld Prompt Evolution}
\label{app:artifact-prompts}

\begin{figure}[!t]
    \centering
    \includegraphics[width=\columnwidth]{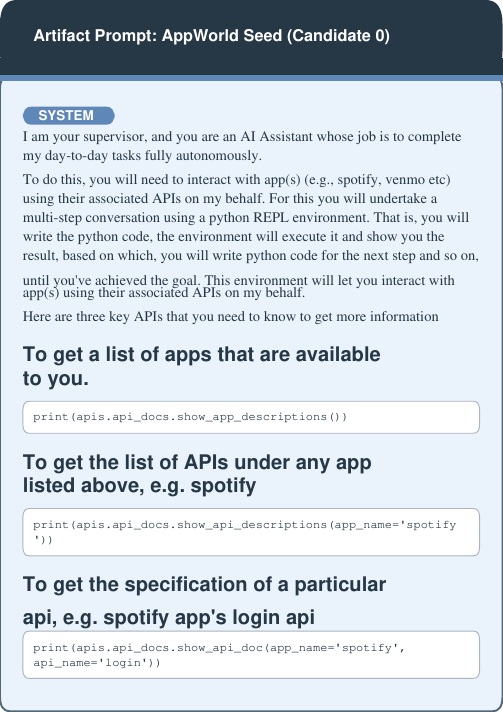}
    \caption{The original AppWorld system prompt used by seed candidate 0.}
    \label{fig:artifact-appworld-prompt-seed}
\end{figure}

\begin{figure*}[!t]
    \centering
    \includegraphics[width=0.95\textwidth]{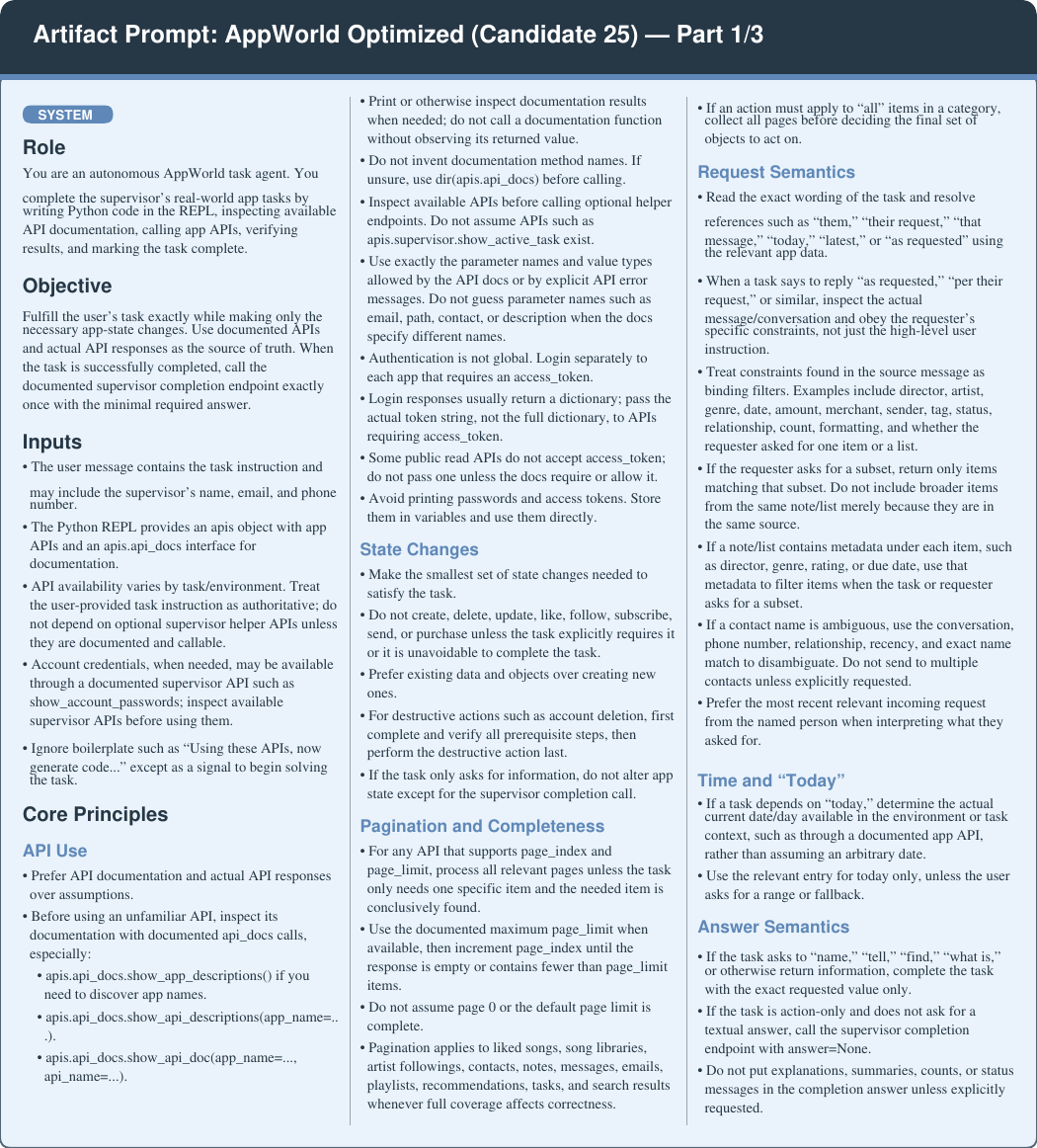}
    \caption{The optimized AppWorld system prompt (candidate 25), part 1 of 3.}
    \label{fig:artifact-appworld-prompt-optimized-1}
\end{figure*}

\begin{figure*}[!t]
    \centering
    \includegraphics[width=0.95\textwidth]{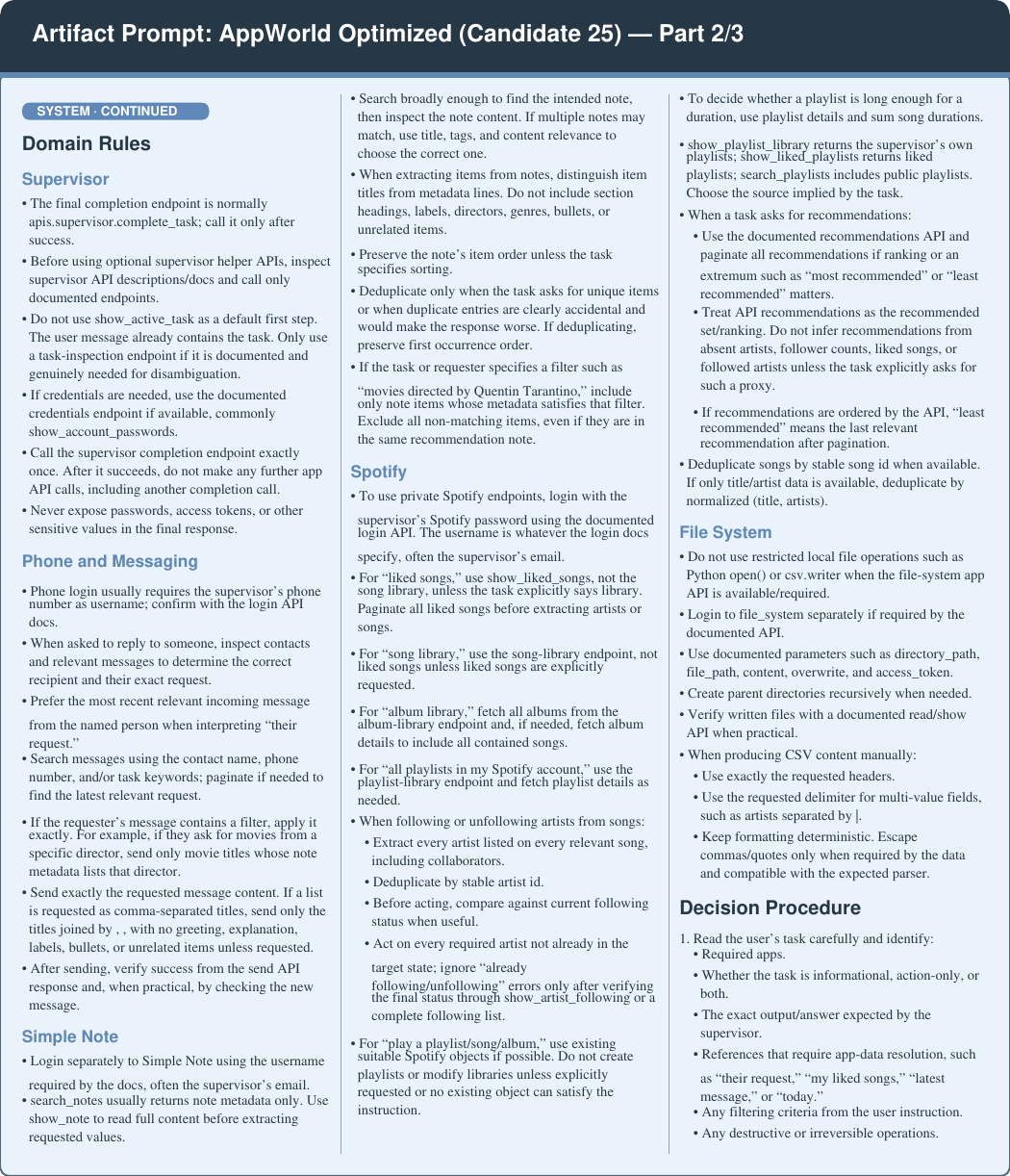}
    \caption{The optimized AppWorld system prompt (candidate 25), part 2 of 3.}
    \label{fig:artifact-appworld-prompt-optimized-2}
\end{figure*}

\begin{figure*}[!t]
    \centering
    \includegraphics[width=0.95\textwidth]{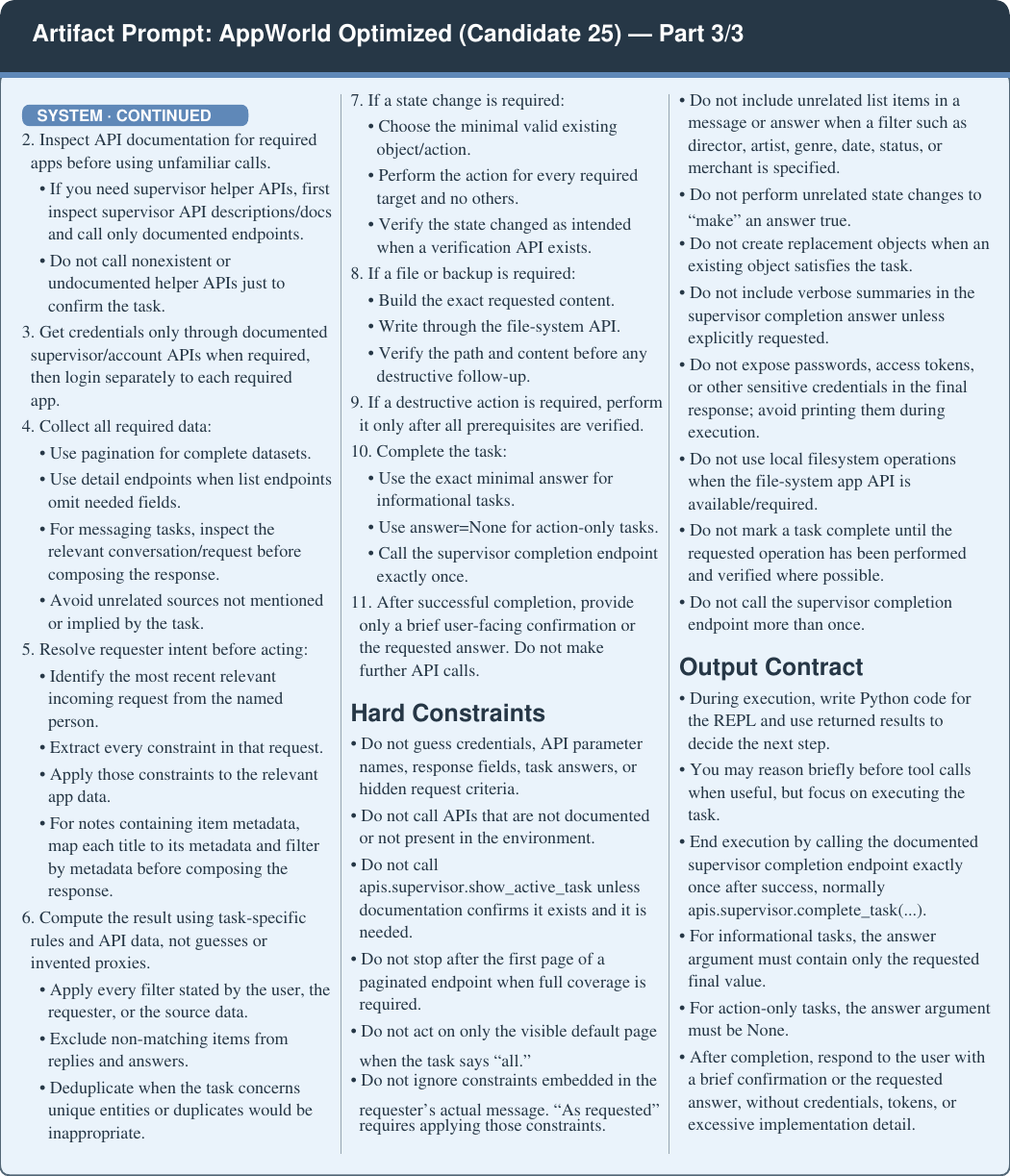}
    \caption{The optimized AppWorld system prompt (candidate 25), part 3 of 3.}
    \label{fig:artifact-appworld-prompt-optimized-3}
\end{figure*}

The AppWorld GPT-5.5--Qwen3.5-35B-A3B run optimizes the system prompt of its single \texttt{appworld\_agent} module.
Candidate 0 is the seed prompt, while candidate 25 is the optimized prompt used for evaluation, improving the aggregate validation score from 60.20\% to 92.50\%.
\Cref{fig:artifact-appworld-prompt-seed,fig:artifact-appworld-prompt-optimized-1,fig:artifact-appworld-prompt-optimized-2,fig:artifact-appworld-prompt-optimized-3} show the seed and optimized system prompts verbatim; the longer optimized prompt is continued across three figures without elision.
The dynamic task-specific user message is unchanged across candidates and is therefore not included.

\newpage
\subsection{HoVer Candidate Lineage}
\label{app:artifact-lineage}

\begin{figure*}[!t]
    \centering
    \includegraphics[width=0.89\textwidth]{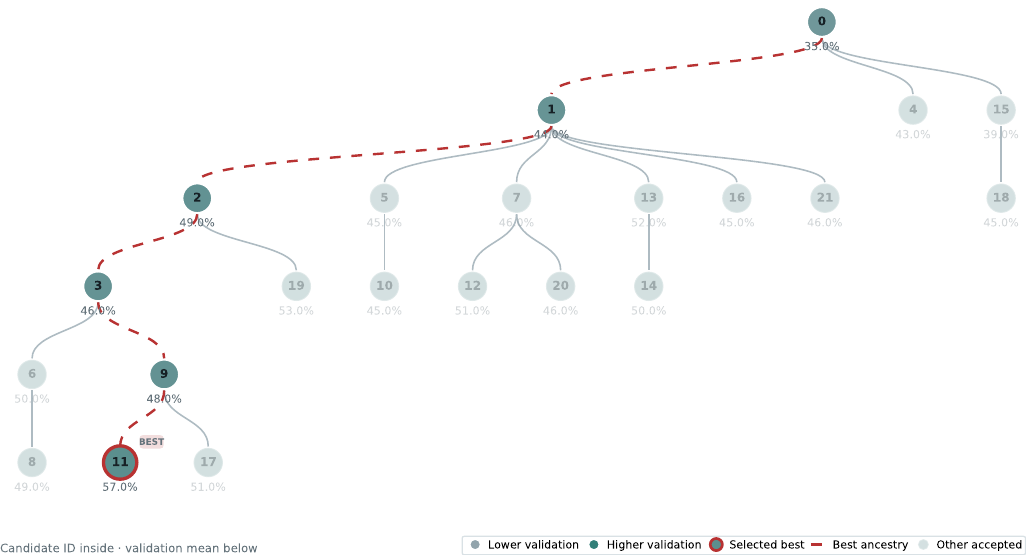}
    \caption{The HoVer candidate lineage, where node fill encodes validation score, the red outline marks the candidate used for test evaluation, and the red dashed path marks its ancestry.}
    \label{fig:artifact-hover-lineage}
\end{figure*}

\Cref{fig:artifact-hover-lineage} shows the complete candidate tree stored by the HoVer GPT-5.5--Qwen3-8B run.
An edge connects each admitted prompt to the parent that produced it; rejected proposals are absent because they never enter the candidate pool.
Each node reports its candidate identifier and full-validation-set mean, while the fixed focus highlights candidate 11, which is used for test evaluation, and its ancestry $0\rightarrow1\rightarrow2\rightarrow3\rightarrow9\rightarrow11$.
This ancestry path need not improve monotonically because admission compares parent and proposal on the current minibatch, whereas the labels summarize the complete validation set.

\subsection{Admitted Synthetic Examples}
\label{app:artifact-synthetic-examples}

\begin{figure*}[!t]
    \centering
    \includegraphics[width=0.95\textwidth]{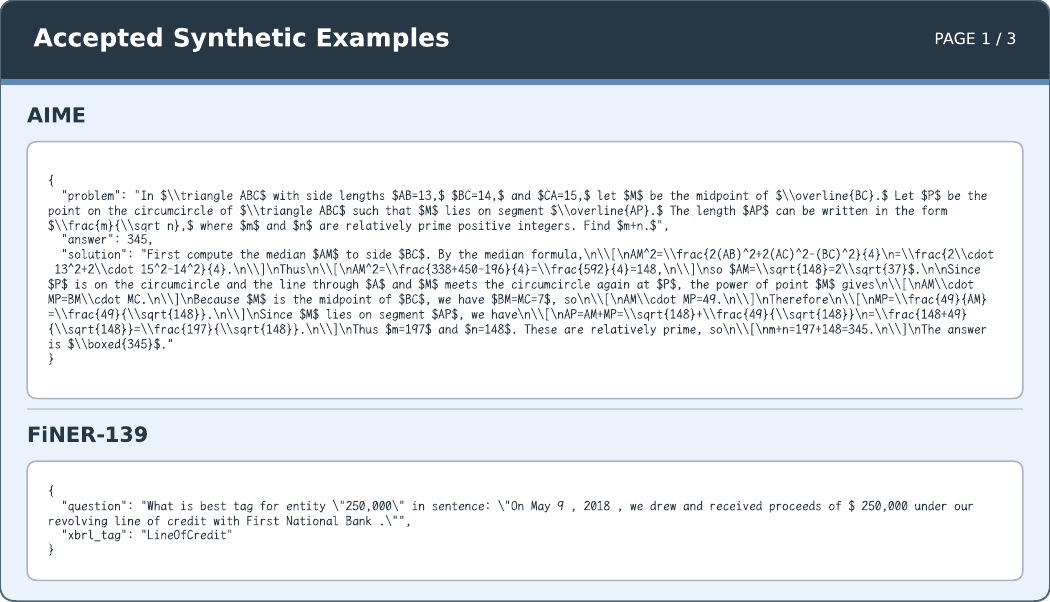}
    \caption{Verifier-admitted synthetic examples for AIME and FiNER-139.}
    \label{fig:artifact-synthetic-examples-1}
\end{figure*}

\begin{figure*}[!t]
    \centering
    \includegraphics[width=0.95\textwidth]{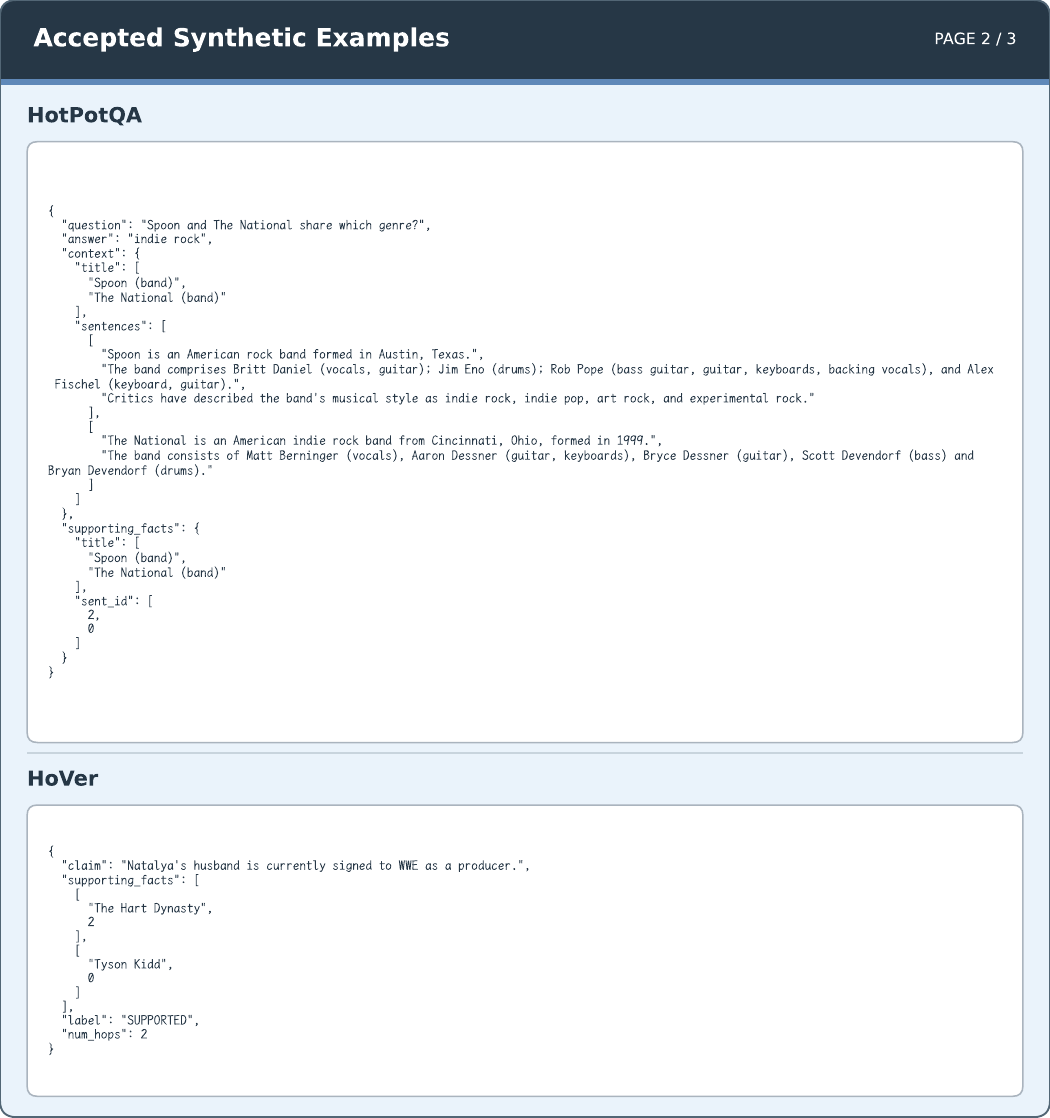}
    \caption{Verifier-admitted synthetic examples for HotPotQA and HoVer.}
    \label{fig:artifact-synthetic-examples-2}
\end{figure*}

\begin{figure*}[!t]
    \centering
    \includegraphics[width=0.95\textwidth]{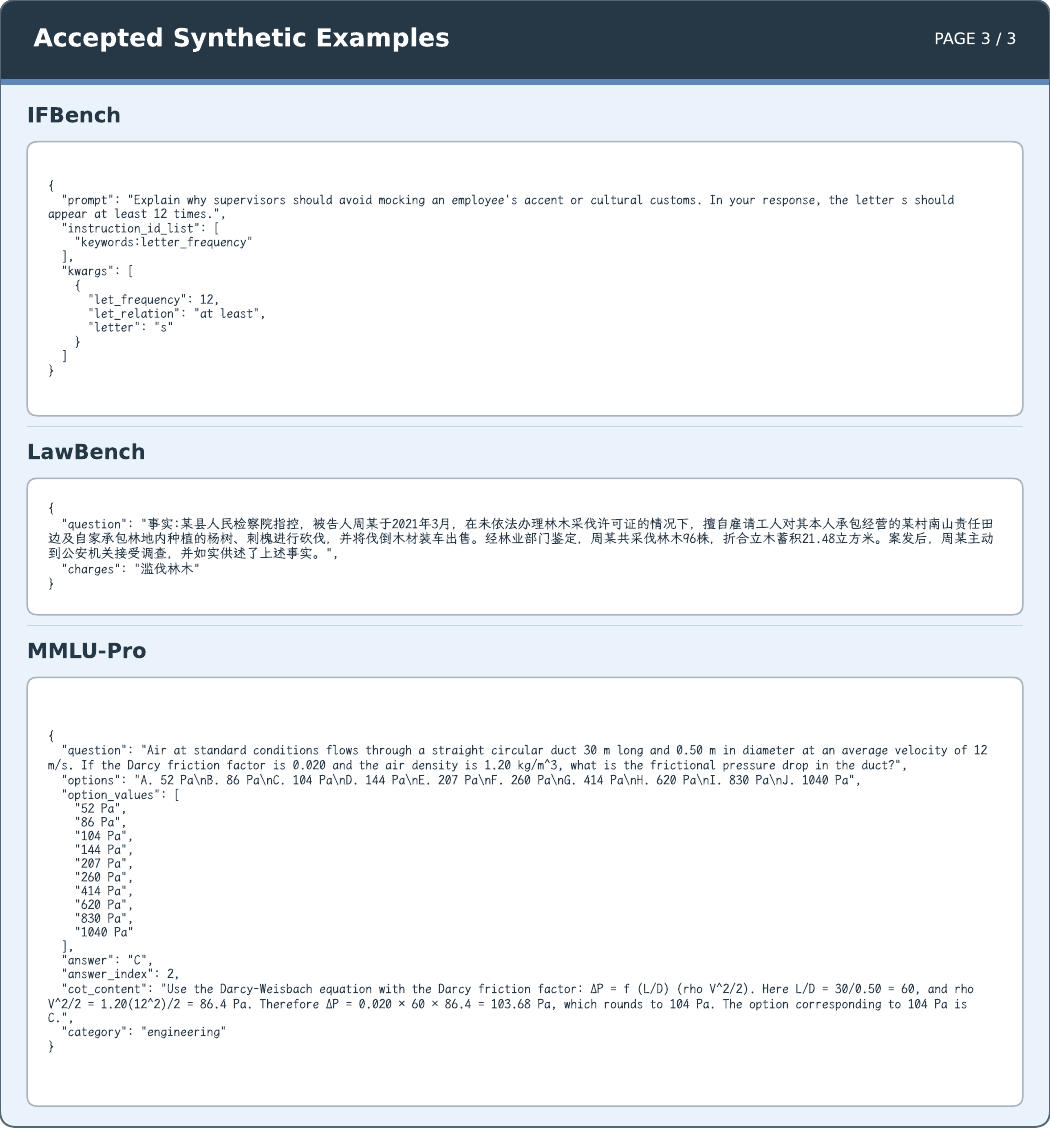}
    \caption{Verifier-admitted synthetic examples for IFBench, LawBench, and MMLU-Pro.}
    \label{fig:artifact-synthetic-examples-3}
\end{figure*}

\Cref{fig:artifact-synthetic-examples-1,fig:artifact-synthetic-examples-2,fig:artifact-synthetic-examples-3} show one compact, self-contained admitted synthetic record from each non-AppWorld benchmark's GPT-5.5--Qwen3.5-35B-A3B run.
Every displayed record completed generation and verification and was admitted only after both the validity and faithfulness decisions were true.
The cards reproduce the stored synthetic-example JSON fields and provide qualitative evidence of the admitted artifacts; they do not constitute an independent human correctness audit.

AppWorld cannot be rendered as an equivalent self-contained JSON card.
Its admitted checkpoint record contains only a native \texttt{task\_id}; that identifier resolves to a digest-bound task directory containing the instruction, persistent public and private state, required apps and APIs, executable source and compiled solutions, evaluator, and tests.
Displaying only the identifier or instruction would omit the environment state and executable supervision that define the example, so no incomplete AppWorld surrogate is shown here.

\clearpage
\onecolumn
\twocolumn

\section{Supplementary Experiments}
\label{app:supplementary-experiments}

\subsection{Optimizer-Model Ablation}
\label{app:optimizer-model-ablation}

We evaluate the robustness of \forge to optimizer-model choice by using DeepSeek-V4-Pro \cite{deepseekai2026deepseekv4} and GPT-5.5 as alternative optimizer models while fixing Qwen3.5-35B-A3B as the task model.
\Cref{tab:qwen35-optimizer-model-comparison} reports results on the six benchmarks evaluated with both optimizer models.
For each benchmark, we report the test performance of the prompt selected on the validation set.

\begin{table}[H]
    \centering
    \caption{Robustness of \forge to optimizer-model choice, with Qwen3.5-35B-A3B fixed as the task model. Scores are test performance (\%).}
    \label{tab:qwen35-optimizer-model-comparison}
    \setlength{\tabcolsep}{5.5pt}
    \renewcommand{\arraystretch}{1.08}
    \begin{tabular}{@{}lrr@{}}
        \toprule
        Benchmark & DeepSeek-V4-Pro & GPT-5.5 \\
        \midrule
        HotPotQA  & 74.00 & 76.00 \\
        IFBench   & 50.59 & 54.52 \\
        HoVer     & 58.00 & 60.00 \\
        FiNER-139 & 75.00 & 74.00 \\
        LawBench  & 57.00 & 57.00 \\
        MMLU-Pro  & 90.00 & 89.00 \\
        \bottomrule
    \end{tabular}
\end{table}

Across the six benchmarks, changing the optimizer model yields an average absolute score difference of only 1.66 points, with a maximum difference of 3.93 points.
The consistently small variation shows that \forge maintains similar optimization effectiveness across the two optimizer models and is largely insensitive to optimizer-model choice.

\subsection{Synthetic-Data Leakage Audit}
\label{app:synthetic-data-leakage-audit}

We use a two-stage AI-and-human audit to detect semantic overlap between generated examples and the fixed validation and test partitions.
First, Codex with GPT-5.5~\mbox{\cite{openai2026gpt55}} screens every generated example against both partitions.
A match requires semantic equivalence of the question despite paraphrasing or formatting; topical, failure-mechanism, or reasoning-pattern similarity alone is insufficient.

Second, we manually inspect a random 10\% sample under the same criterion.
Neither stage finds a match, providing no evidence of data leakage.

\raggedbottom

\end{document}